# Review and modeling of charge carrier mobility in diamond

Faiz Rahman Is-haqzai [a,b,*], Muhammed Deniz [b], Kevin Kröninger [a], Jens Weingarten [a]

[a] AG Kröninger, Faculty of Physics, TU Dortmund University, Otto-Hahn-Str. 4a, Dortmund 44227, Germany
[b] Department of Physics, Dokuz Eylül University, Buca, İzmir TR35160, Turkiye

ABSTRACT

Reported electron and hole mobilities and saturation velocities in diamond vary widely across the literature. We provide a consolidated review of first-principles predicted and experimentally measured mobility and saturation-velocity values in diamond, alongside a focused assessment of the semi-empirical mobility models used to extract low-field mobilities and high-field saturation velocities. We attribute the dispersion primarily to (i) the electric-field window probed in TCT measurements, (ii) the choice of mobility model, and (iii) the excitation source ($\alpha$, laser, or electron). Using an aggregated literature dataset, we benchmark the Trofimenkoff and Caughey–Thomas parameterizations together with a new piecewise model for both conduction- and valence-band transport. For electrons, the piecewise model provides the best global description over a broad electric-field range. It can be interpreted as the room-temperature limit of a more general superposition framework that explicitly incorporates intervalley repopulation in the conduction band. For holes, the Caughey–Thomas model remains the statistically preferred description, consistent with the absence of a repopulation signature in the accessible data. Furthermore, we demonstrate a systematic source dependence ($\alpha$ versus laser) and quantify its impact on fitted mobility and saturation-velocity values. We provide temperature scalings over narrow intervals around room temperature to support Jacoboni–Canali-type parameterization for diamond. Together, these results reconcile much of the apparent inconsistency in the literature and offer guidance for model selection, experimental design, and device-level simulation of charge transport in intrinsic diamond.

## 1. Introduction

The exceptional combination of radiation hardness, temperature resistance, chemical inertness, wide bandgap, and high charge-carrier mobility and saturation velocity makes diamond an ideal candidate for a sensor material for operation under high-rate, high-temperature, and high-radiation conditions. These attributes have motivated its use in high-energy physics (HEP) [1–3], nuclear engineering [4–6], fusion energy [7–9], space applications [10], and dark-matter detection [11]. Owing to its near-tissue equivalence in terms of atomic number and stopping power, diamond is also employed as a sensor material in radiation therapy (RT), including external beam radiotherapy (EBRT) [12–17], intraoperative radiotherapy (IORT) [18–20], soft X-rays and MeV electrons [21,22], and FLASH radiotherapy (FLASH-RT) [23].

Beyond its properties as a sensor material, diamond is listed among the "emerging semiconductor substrates" for power electronics and radio-frequency (RF) applications [24,25]. The future mass production of electronic devices will require environmentally friendly semiconductor materials. In Europe, for example, lists of hazardous substances have been established under the REACH (Registration, Evaluation, Authorization and Restriction of Chemicals) regulation [26]. This makes diamond one of the best candidates for future electronic substrates [27], thereby helping to create more efficient and cost-effective radiation detectors.

This work reviews some of diamond's key charge-carrier properties that are crucial both for its use as a sensor material and as an electronic substrate. These properties affect charge-collection efficiency, signal amplitude, and the energy and timing resolution of diamond radiation detectors. They also critically influence performance parameters such as breakdown voltage, power density, and operating frequency in high-power, high-frequency diamond electronic devices. The principal focus is on the low-field mobility $\mu_0$, effective mobility $\mu_{eff}$, drift velocity $v_d$, and saturation velocity $v_{sat}$ as functions of electric field, temperature, and impurity concentration, for which reported values vary widely in the literature. The low-field mobility, $\mu_0$, and the saturation velocity, $v_{sat}$, have been reported in the ranges $\mu_{0,e}$ = 1300–4500 $cm^2\,V^{-1}\,s^{-1}$, $\mu_{0,h}$ = 2000–3800 $cm^2\,V^{-1}\,s^{-1}$, and $v_{s,e}$ = 9–26.3 × $10^6$ $cm\,s^{-1}$, $v_{s,h}$ = 10–15.7 × $10^6$ $cm\,s^{-1}$. This paper investigates the origins of this spread

* Corresponding author at: AG Kröninger, Faculty of Physics, TU Dortmund University, Otto-Hahn-Str. 4a, Dortmund 44227, Germany.
*E-mail address:* faiz.ishaqzai@tu-dortmund.de (F.R. Is-haqzai).

**List of abbreviations**

| | |
|---|---|
| AIC | Akaike information criterion |
| API | Acoustic-phonon interactions |
| BIC | Bayesian information criterion |
| BTE | Boltzmann transport equation |
| CCE | Charge-collection efficiency |
| CCD | Charge-collection distance |
| CT | Caughey–Thomas Model |
| CVD | Chemical vapor deposition |
| DDM | Drift–diffusion model |
| DM | Drude model |
| DO | Detector Operational (electric-field) range |
| HMT | Hall measurement technique |
| MIM | Metal–insulator–metal |
| OPI | Optical-phonon interactions |
| pcCVD | Polycrystalline chemical vapor deposition |
| PW | Piecewise (mobility model) |
| SCFC | Space-charge-free current |
| scCVD | Single-crystal chemical vapor deposition |
| SCLC | Space-charge-limited current |
| TCT | Transient current technique |
| TK | Trofimenkoff model |
| ToF | Time of flight |
| TRCR | Time-resolved cyclotron resonance |

by comparing sample qualities; experimental techniques, setups and conditions; and data-analysis approaches (models) that may account for the differences. We also review the semi empirical models used to extract the low field mobilities and high field saturation velocities. These reviewed models are optimized and parameterized for diamond's mobility dependence on electric field, temperature, impurity concentration, and carrier–carrier scattering; consistent with the major scattering mechanisms governing the charge carrier transport in semiconductors (see Section 3.1). Because the literature employs different transport models (for electric field dependent mobility), we aggregate and standardize published measurements to evaluate candidate models within a common statistical framework. Model selection is based on $\chi^2/\mathrm{ndf}$ for goodness-of-fit and information criteria (AIC/BIC), which penalize model complexity, with supplementary assessment via $R^2$ and standardized residual z-score metrics ($\sigma_z$) (see Section 5.2). The resulting parameterization will serve as a reference for device-level simulation over the relevant temperature and electric-field ranges.

Section 2 shortly summarizes the historical development of diamond detectors as a motivating context for the review and modeling in this work. Section 3 introduces charge-carrier transport in diamond, emphasizes the central role of mobility in charge carrier transport, reports the first-principles mobility values, and reviews the semi-empirical mobility models commonly used in detector and device simulation with optimization for diamond detectors. Section 4 surveys experimental approaches to mobility extraction, with emphasis on the transient current technique (TCT), and discusses practical issues such as transport regime, polarization, and transientt-time definitions. Reported low-field mobilities and saturation velocities are compiled and compared in Section 4.4. Section 5 benchmarks candidate parametrizations on a common, digitized literature dataset—both experiment-wise within limited field windows and via pooled fits across datasets to constrain the full field dependence—while quantifying excitation-source-dependent normalization offsets through a simple cross-dataset scaling. Section 6 addresses the temperature dependence of mobility, quantifies repopulation effect (electrons) as a superposition framework. It is also important because it provides explanation why different models describe transport data for electrons and holes at room temperature. Section 6.3 provides near-room-temperature scaling of TK/CT parameters as a step toward Jacoboni–Canali-type modeling for diamond. Finally, Section 7 summarizes the main conclusions, gives practical recommendations for simulation tools, and outlines priorities for future experimental campaigns. Recommended values for simulators are provided in Table 11.

## 2. Historical development of diamond detectors

In 1923, diamond was used for the first time as a detector material when a photoconductive UV diamond detector was employed to investigate the absorption characteristics of natural diamonds [28]. This was followed by the design of solid-state ionization detectors—so-called "crystal counters"—to replace traditional gas counters in experimental nuclear physics [29]. A comprehensive history of the development of diamond detectors prior to synthetic diamond is provided in Ref. [30]. In the late 1970s and early 1980s, a series of improved transient-current technique (TCT) experiments confirmed the higher charge-carrier mobilities and saturation velocities in natural diamond relative to silicon [31–34]. However, the use of natural diamond detectors was restricted by their limited availability and uncontrolled impurities. Synthetic diamond technologies were developed to address these problems. Currently, high-pressure, high-temperature (HPHT) and chemical vapor deposition (CVD) are two established technologies for producing synthetic diamond [35,36].

Soon after the establishment of CVD technology, beam tests were conducted to assess the feasibility of CVD diamond radiation detectors in the high-rate, high-radiation environments of the then-proposed SSC and LHC experiments [37]. Subsequently, a solid-state ionization-chamber–type CVD diamond detector in a metal–insulator–metal (MIM) configuration was proposed in 1993 [30]. In 2004, the first polycrystalline CVD (pcCVD) diamond detector was deployed in the BaBar experiment at PEP-II as part of the radiation-monitoring system [1]. Since then, pcCVD diamonds have been used as beam-condition monitors (BCMs) and/or beam-loss monitors (BLMs) not only in BaBar [3] but also in other HEP experiments—e.g., ATLAS [38,39], CMS [40], and LHCb [41] at CERN—and at the Collider Detector at Fermilab (CDF) [42]. The High-Luminosity upgrade of the Large Hadron Collider (HL-LHC) plans to utilize pcCVD diamonds as BCMs [43]. The Belle II experiment at the SuperKEKB asymmetric-energy $e^+e^-$ collider uses single-crystal CVD (scCVD) diamonds as BLMs [44], representing the first deployment of an scCVD diamond sensor as a BLM.

## 3. Charge carrier transport and mobility modeling in diamond

Charge transport in diamond can be described microscopically by the Boltzmann transport equation, but device-level simulations typically rely on reduced models such as drift–diffusion, in which the mobility acts as the key material parameter linking microscopic scattering to macroscopic current flow. This section therefore (i) introduces the transport framework and the scattering mechanisms that govern mobility in diamond, (ii) compiles and discusses reported low-field mobilities from first-principles and related theoretical studies of diamond, as summarized in Table 1, and (iii) summarizes semi-empirical mobility models used in detector and device simulation, emphasizing how widely used parameterizations (field, temperature, doping, and carrier–carrier scattering) were developed and subsequently adapted for diamond.

### 3.1. Transport framework and mobility as an emergent parameter

The semi-classical Boltzmann transport equation (BTE) provides the most general description of charge-carrier dynamics in crystalline solids, including semiconductors. It tracks the evolution of the charge-carrier distribution function $f(\vec{r},\vec{k},t)$ in phase space under external fields and scattering, where $\vec{r}$ is position, $\vec{k}$ is the Bloch wave vector, and $t$ is time. However, solving the full BTE is computationally prohibitive for device-scale simulations, therefore, simplified macroscopic transport models are employed to obtain tractable equations, such as the drift–diffusion model (DDM) [45,46].

When external perturbations are modest so that charge carriers remain close to local equilibrium, scattering drives the distribution toward a displaced Maxwell–Boltzmann (or Fermi–Dirac) form on a characteristic relaxation time $\tau$. Linearizing the collision integral in this relaxation-time approximation (RTA) and taking velocity moments produces continuity equations coupled to current-density relations. For non-degenerate semiconductors, this procedure leads to the familiar DDM,

$$\vec{J}_i = q\,\mu_i\,n_i\,\vec{E} \;+\; \mu_i\,k_B\,T\,\vec{\nabla}n_i, \tag{1}$$

where $i \in \{n,p\}$ and $\mu_i$ denotes the carrier mobility. Note that $\mu_i$ governs not only the drift term but also diffusion through the Einstein relation $D_i = \mu_i k_B T/q$. This model is widely used in radiation-detector simulation frameworks to calculate signal-related parameters such as signal amplitude, charge-collection efficiency (CCE), and charge-collection distance (CCD) [47].

In the spatially homogeneous limit ($\vec{\nabla}n_i = 0$), Eq. (1) reduces to the Drude model (Eq. (2)), $\vec{J}_i = \sigma_i\vec{E}$, with conductivity $\sigma_i = q\,\mu_i\,n_i$. The historical Drude model [48], first proposed for metals, treats carriers as classical particles accelerated by the Lorentz force $q\vec{E}$ between randomizing scattering events that occur, on average, after a time $\tau$ (the relaxation time). Momentum balance then yields a steady-state drift velocity,

$$v_d = \frac{q\,\tau}{m^*}E = \mu E \;\Rightarrow\; \frac{v_d}{E} = \mu \equiv \frac{q\,\tau}{m^*}, \tag{2}$$

where $m^*$ is the effective mass of the charge carriers. Although oversimplified for modern semiconductors, the Drude picture remains valuable because it defines mobility as the key material parameter linking microscopic scattering physics to macroscopic current flow. Eq. (2) makes clear that $v_d$ and $\mu$ depend on the scattering rate $1/\tau$. These scattering mechanisms are broadly classified into (a) lattice scattering, (b) defect scattering, and (c) carrier–carrier scattering.

Lattice scattering encompasses acoustic-phonon interactions (API), optical-phonon interactions (OPI), and intervalley scattering. These interactions can be either activated or suppressed by external conditions such as the electric field and temperature. In this work, we focus on lattice scattering (API, OPI, intervalley), for which the dependence on electric field and temperature aligns with sensor operating conditions. The literature provides a relatively rich dataset for review and further analysis.

Defect scattering is typically classified into (a) crystal defects, (b) impurities, and (c) alloy scattering. Usable data exist in the diamond literature about impurity scattering and we will cover it in a separate paper. The effects of the remaining scattering mechanisms on diamond mobility remain to be investigated.

### 3.2. Predicted mobility values in diamond

For device-scale simulations, models such as the DDM (Eq. (1)) require a mobility input that can be estimated experimentally (see Section 4) or computed from microscopic theory. For theoretical computations, one typically solves the momentum-space (k − space) BTE using material-specific electronic band structures, phonon spectra, and electron–phonon scattering matrix elements obtained from first-principles electronic-structure calculations for bulk, homogeneous systems.

A broad range of first-principles and quasi-first-principles mobility codes has emerged over the past two decades. A comprehensive overview of these strategies and codes is given in Ref. [49]. Despite this progress, comparatively few studies have applied these methods to compute charge-carrier mobility in diamond.

Perhaps the earliest application of the BTE in the relaxation-time approximation (BTE–RTA) to estimate mobility in diamond dates to 1948 [50], when Seitz predicted the low-field electron mobility to be $\mu_{0,e} \approx 156$ cm$^2$ V$^{-1}$ s$^{-1}$.

**Table 1**
Summary of theoretical charge-carrier mobility studies in diamond.

| Ref. | $\mu_{0,e/h}$ (cm$^2$/Vs) | Comments |
|---|---|---|
| [50] | 156/– | Analytical perturbation-theory model for acoustic-phonon interaction (API) at 300 K. |
| [51] | –/**1830** | Three-band, low-field transport model, phonon-limited at 300 K for $N_A \lesssim 10^{17}$ cm$^{-3}$. |
| [52] | 130/– | First-principles electron–phonon (Elliott–Yafet) calculation at 300 K. |
| [53] | –/**2500** | Magnetotransport, phonon-limited (PL) at 300 K (doped). |
| [54] | 1790/**1970** | PL mobility versus temperature obtained by solving the BTE. |
| [55] | 2400/**2340** | $\mu$ along [100]. Full-band Monte Carlo model (FBMCM) for PL transport. |
| [55] | –/**1830** | $\mu$ along [110]. FBMCM for PL transport. |

*Note:* Bold numbers denote hole mobilities, a double dash (–) indicates that the carrier type was not treated in that study.

Some recent works include a study of p-doped diamond in 2010 [51] that considered four scattering mechanisms in a semi-empirical model, neutral and ionized impurity scattering, and acoustic and nonpolar optical phonon scattering. The intrinsic hole mobility was found to be 1830 cm$^2$ V$^{-1}$ s$^{-1}$. In 2012 [52], a first-principles study using the electronic spin-relaxation rate of diamond reported an electron mobility of 130 cm$^2$ V$^{-1}$ s$^{-1}$ at room temperature. In their 2018 study, Macheda and Bonini solved the ab initio BTE including the effect of a finite magnetic field and determined the drift mobility, Hall mobility, and Hall factor. The calculated hole mobility at room temperature was 2500 cm$^2$ V$^{-1}$ s$^{-1}$ [53]. Table 1 summarizes some of the theoretically calculated mobilities for electrons and holes obtained using different approaches and approximations.

Theoretical frameworks are also used to identify the dominant scattering mechanisms over specific electric-field and/or temperature ranges for characteristic impurity concentrations. Such analyses help refine semi-empirical mobility models by isolating the dominant and competing mechanisms from experimentally obtained $v_d$–$E$, $v_d$–$T$, and $v_d$–$N$ characteristics.

### 3.3. Semi-empirical mobility models

As mentioned in Section 3.2, charge carrier mobility in diamond can be calculated from first principles, but this requires a tremendous amount of computational power, and the predicted values often do not agree with measurements and cannot be used reliably in device-level simulation frameworks. For this reason, semi-empirical mobility models have been developed. They are fitted to experimental data and used to extract parameter values, including low-field mobilities in semiconductors. In this section, we review the development of these semi-empirical mobility models, from the simple Drude model (Eq. (34)) to formulations that incorporate the effects of high electric fields, temperature, impurity concentration, and carrier–carrier scattering. The specific focus is on the review of experimental reparametrization of these models for use in diamond.

#### 3.3.1. Mobility vs electric field

If $v_d$ does not saturate, the Drude model (Eq. (2)) can be used to estimate the low-field mobility $\mu_0$. However, this model fails to describe the data when $v_d$ is characterized above $E_c$ and begins to scale nonlinearly with E, i.e., $v_d = \mu(E)E$ (see Section 4.3). The Drude model (Eq. (2)) has therefore been empirically extended to describe experimental data over wide electric-field ranges.

Dacey and Rose extended the Drude model (Eq. (2)) by introducing the concept of a critical field $E_c$ for the first time (see Section 4.3) [56],

$$v_d = v_s \frac{E}{E_c}\left(\frac{E_c}{E}\right)^{\frac{1}{2}}, \quad (3)$$

which can also be written as $\mu_{eff} = (\mu_0/E)(E_c/E)^{1/2}$ following Eq. (21), relating the effective mobility $\mu_{eff}$ to the constant low-field mobility $\mu_0$, where $\mu_0 = v_s/E_c$ and $\mu_{eff} = v_d/E$ (see Section 4.3). $v_s$ is the velocity at $E > E_c$. All later models follow the same principle and can be expressed either in terms of $v_d$ or $\mu_{eff}$. Trofimenkoff [57] further improved the Dacey–Rose model (Eq. (3)), which is widely used for characterizing charge-carrier mobilities in diamond,

$$v_d = v_s \frac{E/E_c}{1 + E/E_c}. \quad (4)$$

A shaping parameter $\beta$ was later added to the Trofimenkoff (TK) model (Eq. (4)) [58] to obtain the widely used Caughey–Thomas (CT) model in silicon,

$$v_d = v_s \frac{E/E_c}{\left(1 + (E/E_c)^{\beta}\right)^{1/\beta}}. \quad (5)$$

This evolution reflects a series of empirical extensions of the theoretical framework founded by Drude to describe experimental data.

A high-electric-field model [59] has been developed to better describe the $v_{d,h}$ behavior at very high electric fields in intrinsic silicon. A similar extension is needed to model $v_d(E)$ at very high fields in diamond, because the CT model (Eq. (5)) does not fit well in the high-electric-field regime (see Fig. 3). We address this in Sections 5 and 6.

### 3.3.2. Mobility vs temperature

Low-field mobility $\mu_0$ and other parameters ($E_c$, $v_s$, $\beta$) in the TK and CT models scale with temperature $T$ using the relation,

$$P(T) = P_{ref}\left(\frac{T}{T_{ref}}\right)^{-\alpha} = P_{ref}\, T_{ref}^{\alpha}\, T^{-\alpha} = A_{ref}\, T^{\gamma}. \quad (6)$$

Here, $P$ is a temperature-dependent model parameter and $P_{ref} \equiv P(T_{ref})$. The expressions are equivalent upon setting $A_{ref} = P_{ref} T_{ref}^{\alpha}$ and $\gamma = -\alpha$. Jacoboni et al. [60] extended the original CT model by introducing temperature scaling, yielding the $v_d(E,T) = \mu_{eff}(E,T)E$ model form for silicon. The same approach is used for diamond [31,61,62]. The interpretation of temperature-dependent Hall mobility measurements in diamond also follow this approach [63–68].

Quay et al. [69] developed a temperature-scaling mobility model for several semiconductors, needed to be parameterized for diamond.

### 3.3.3. Mobility vs impurity concentrations

Caughey and Thomas [58] described quantitatively, for the first time in silicon, the doping dependence of carrier mobility,

$$\mu_{eff} = \frac{\mu_{max} - \mu_{min}}{1 + (N/N_{ref})^{\alpha}} + \mu_{min}. \quad (7)$$

This model was extended by Arora et al. [70] to include temperature, i.e., $v_d(N,T) = \mu_{eff}(N,T)E$, and by Masetti et al. [71] to Ar-, P-, and B-doped silicon,

$$\mu_e(N) = \frac{\mu_{max,e} - \mu_{0,e}}{1 + (N/C_{r,e})^{\alpha_e}} + \mu_{0,e} - \frac{\mu_{1,e}}{1 + (C_{s,e}/N)^{\beta_e}}, \quad (8)$$

$$\mu_h(N) = \frac{\mu_{max,h}}{1 + (N/C_{r,h})^{\alpha_h}} + \frac{\mu_{0,h}}{e^{P_c/N}} - \frac{\mu_{1,h}}{1 + (C_{s,h}/N)^{\beta_h}}, \quad (9)$$

where $\alpha$ and $\beta$ are dimensionless shaping parameters and $C_r$ is a reference dopant concentration, analogous to $N_{ref}$ in Eq. (7).

The Masetti model (Eq. (9)) was parameterized for B-doped diamond by Rashid [72]. He also proposed a model $\mu_h(T, N_A) = v_{d,h}(T, N_A)E$ that incorporates the temperature dependence alongside boron acceptor concentration in diamond,

$$\mu_h(T, N_A) = \mu_{min}(T) + \frac{\mu_{max}(T) - \mu_{min}(T)}{1 + (T/300)^{m_1}(N_A/C_0)^{\gamma_2}}, \quad (10)$$

$$\mu_{min/max}(T) = \mu_{min/max,0}(T/300)^{n_{min/max}}, \quad (11)$$

where $\mu_{min/max,0}$, $n_{min/max}$, $m_1$, $C_0$, and $\gamma_2$ are material-specific parameters given in [72].

Kagamihara et al. [73] proposed a model to describe electron transport in SiC,

$$\mu(T, N_{imp}) = \mu(300, N_{imp})\,(T/300)^{-\beta(N_{imp})}, \quad (12)$$

with

$$\mu(300, N_{imp}) = \mu^{min} + \frac{\mu^{max} - \mu^{min}}{1 + (N_{imp}/N_{\mu})^{\gamma_{\mu}}}, \quad (13)$$

and

$$\beta(N_{imp}) = \beta^{min} + \frac{\beta^{max} - \beta^{min}}{1 + (N_{imp}/N_{\beta})^{\gamma_{\beta}}}. \quad (14)$$

Pernot and Koizumi [74] parameterized the Kagamihara model (Eq. (12)) for phosphorus-doped [111] homoepitaxial diamond, while Volpe [75] parameterized it for B-doped diamond.

### 3.3.4. Mobility vs carrier–carrier scattering

Klaassen [76] integrated carrier–carrier (electron–hole) scattering into a unified, physically based analytical framework. This model emphasizes the crucial impact of electron–hole scattering under high-injection (bipolar) conditions, i.e., when $\Delta n = \Delta p$ is comparable to or exceeds the background doping. The mobility $\mu(T, N, cc)$ is calculated using Matthiessen's rule,

$$\frac{1}{\mu(T,N,cc)} = \frac{1}{\mu(T)} + \frac{1}{\mu(N)} + \frac{1}{\mu(cc)}, \quad (15)$$

where $\mu(T)$ is parameterized using Eq. (6). The term $\mu(N)$ is described by a modified Brooks–Herring–type formulation [77],

$$\mu(T, N) = \mu_{min} + \frac{\mu_{max} - \mu_{min}}{1 + (N/N_{ref})^{\alpha}} \left(\frac{T}{300}\right)^{3\alpha - 1.5}. \quad (16)$$

Carrier–carrier scattering mobility $\mu(cc)$ is explicitly modeled for high-injection conditions and written as

$$\mu(cc) = F(P)\,\mu(T, N), \quad (17)$$

$$\text{where} \quad F(P) = \frac{\frac{m_1}{m_2} P^{r_1 + r_2 + r_3}}{\frac{m_1}{m_2} P^{r_1 + r_4 + r_5}}. \quad (18)$$

Here, $\mu_{min/max}$, $N_{ref}$, $\alpha$, and $r_i$ are fitting parameters based on experimental data, $P$ is a dimensionless high-injection parameter and $m_1/m_2$ is an effective-mass ratio.

Klaassen's compensation-aware formulation[1] ensures accuracy in heavily doped silicon, smoothly transitioning between majority- and minority-carrier mobilities, and providing physical fidelity critical for semiconductor device simulation. This model is not yet parameterized for diamond.

An independent work [78] proposed the carrier–carrier scattering mobility in diamond as

$$\mu_{cc} = \frac{3.11 \times 10^{16}\left(T^2/n^{3/2}\right)}{\ln\left(1 + 1.77 \times 10^{8}\left(T^2/n^{2/3}\right)\right)}, \quad (19)$$

where $n$ is the carrier concentration.

[1] Mobility is modeled versus $N_D$ and $N_A$ separately, because both types of ionized impurities act as scattering centers. This yields additional mobility degradation at the same net doping when the material is compensated.

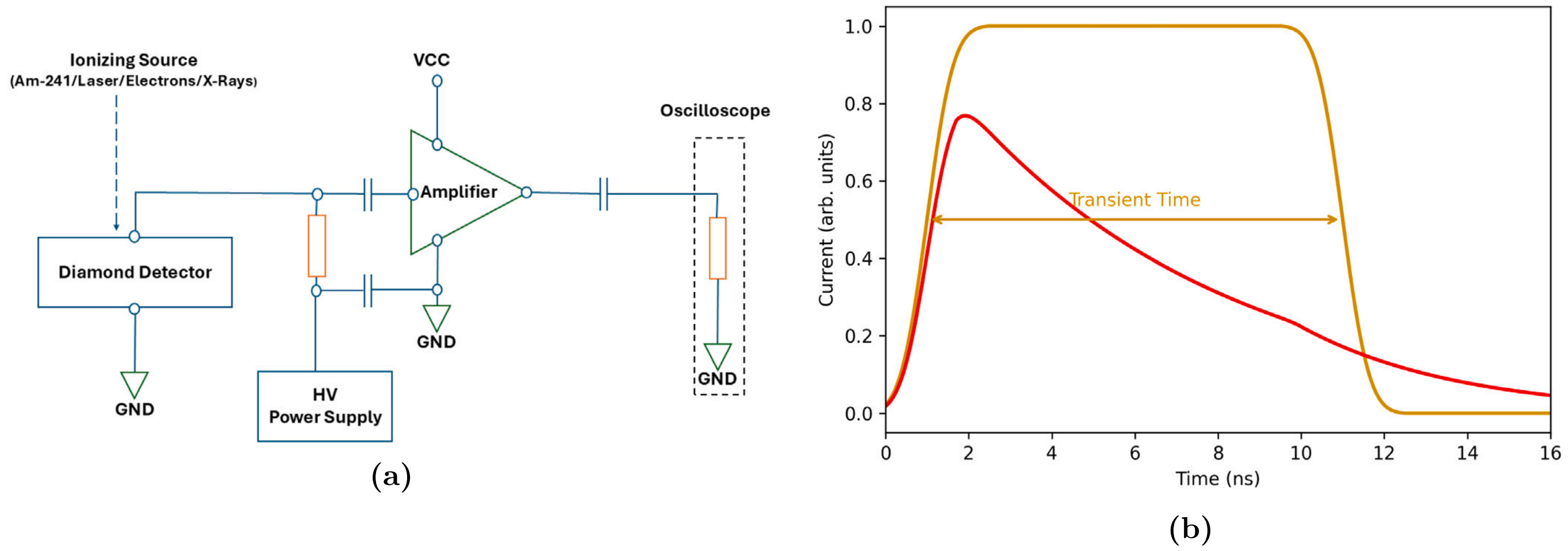

**Fig. 1.** (a) Schematic of a TCT setup used to measure charge carrier mobility in diamond. (b) Idealized TCT waveforms showing the transient current response. The gold trace corresponds to complete charge collection. The carrier transit time is extracted from the full width at half maximum (FWHM). The red trace corresponds to incomplete drift, where the charge cloud does not reach the collecting electrode and the trailing edge exhibits no cusp.

## 4. Experimental considerations and measured mobility values

One way to determine the charge-carrier mobility is to characterize the material directly in the laboratory. Experimentally measured mobilities are essential both for diamond detector engineering and for validating transport models, because they reveal how scattering, trapping, and field distortion manifest under realistic operating conditions. This section summarizes the principal mobility measurement techniques used in diamond and then focuses on the transient current technique (TCT), which is most widely used to obtain $\mu(E,T)$ in detector structures. We also discuss practical issues that govern the reliability of mobility extraction—space-charge-free versus space-charge-limited transport, polarization, and transit-time definitions—before compiling reported $\mu_0$ and $v_s$ values in Section 4.4.

### *4.1. Overview of mobility measurement techniques*

Hall-effect measurement technique (HMT), charge-collection efficiency/distance (CCE/CCD) (see Section 4.4), electron-beam–induced current (EBIC) [79–81], ion-beam–induced charge (IBIC) [82], time-resolved cyclotron resonance (TRCR) [83] (see Section 6), and the transient current technique (TCT) are among the experimental procedures used for mobility measurements.

In the HMT, a dc current I flows along the sample (x-direction) while a magnetic field $\vec{B}$ is applied perpendicular to the sample plane (z-direction). Carriers of the elementary charge $q$ (±e) drift with velocity $\vec{v}_d$ and experience the Lorentz force $q\,\vec{v}_d \times \vec{B}$. In steady state a transverse Hall field $\vec{E}_H$ develops such that

$$q\left(\vec{E}_H + \vec{v}_d \times \vec{B}\right) = 0.$$

The transverse Hall voltage $V_H$ is measured between side contacts separated by the width $w$ (van der Pauw or Hall-bar geometry). This technique was used to determine diamond mobilities in early studies [63–68,84]. They rely on

$$\mu = \frac{\mu_H}{r_H} \approx \mu_H = |R_H|\,\sigma = \frac{\sigma}{qn}, \tag{20}$$

where

$$R_H = \frac{V_H t}{I B}, \qquad \sigma = \frac{1}{\rho} = \frac{I}{V_x}\,\frac{L}{wt},$$

to extract charge-carrier mobilities. Here, $\sigma$ is the conductivity, $\rho$ the resistivity, $R_H$ and $r_H$ are the Hall coefficient and Hall factor, $n$ the carrier concentration, $t$ the sample thickness, and $V_x$ the longitudinal voltage drop over length $L$. For high-purity diamond, $r_H \approx 1$, so the drift mobility satisfies $\mu \approx \mu_{\rm H}$ (the Hall mobility). However, because of the extremely low free-carrier density, the resulting Hall voltage is often very small and susceptible to noise and contact offsets. Nowadays, Hall measurements are used mainly for impurity-dependent mobility characterization.

EBIC has historical importance in diamond. Early work employed a pulsed, TCT-like EBIC implementation combined with Hecht (CCE/CCD) analysis to extract $\mu$, $\tau$, and space-charge/trapping parameters [79,80].

TRCR is a state-of-the-art method (see Section 6) for evaluating mobilities at very low temperatures ($T \lesssim 40$ K). TCT is the most robust method used by experimentalists to extract electric-field- and temperature-dependent mobilities in diamond, accordingly, we discuss it in Section 4.2.

### *4.2. Transient current technique (TCT) for mobility extraction*

TCT measurements of the mobility are typically carried out using planar ionization-chamber sensors in a metal–insulator–metal (MIM) configuration [30]. A high-voltage (HV) power supply is used to apply the electric field across the sensor. The readout system contains a single- or multistage amplifier and an oscilloscope, as shown in Fig. 1(a). A source of ionizing radiation generates charge carriers that drift (electrons or holes) under the applied bias across the sensor thickness, giving rise to a transient current signal (the TCT signal). The full width at half maximum (FWHM) is called the transit time $t_t$ which is the time the generated charge carriers need to traverse a distance d (typically the detector thickness) under the applied electric field E. An example TCT signal is shown in Fig. 1(b). The effective mobility $\mu_{\rm eff}$ can be extracted using the Drude model (Eq. (2)).

$$\mu_{eff} = \frac{v_d}{E} = \frac{d}{t_t E}. \tag{21}$$

Modern scanning-TCT setups use a laser as carrier-generation source. This enables localized charge-carrier generation around the laser's focal point, allowing precise probing of drift near device edges and at arbitrary depths within the bulk [67].

In general, $\alpha$ particles [62,85–92], pulsed electron beams [31–34,93], pulsed X-rays [94,95], or pulsed ultraviolet lasers [61,96–100] are used as ionizing sources. In Section 5, we group literature-reported mobilities by ionization source to directly compare possible source-dependent variations in mobility.

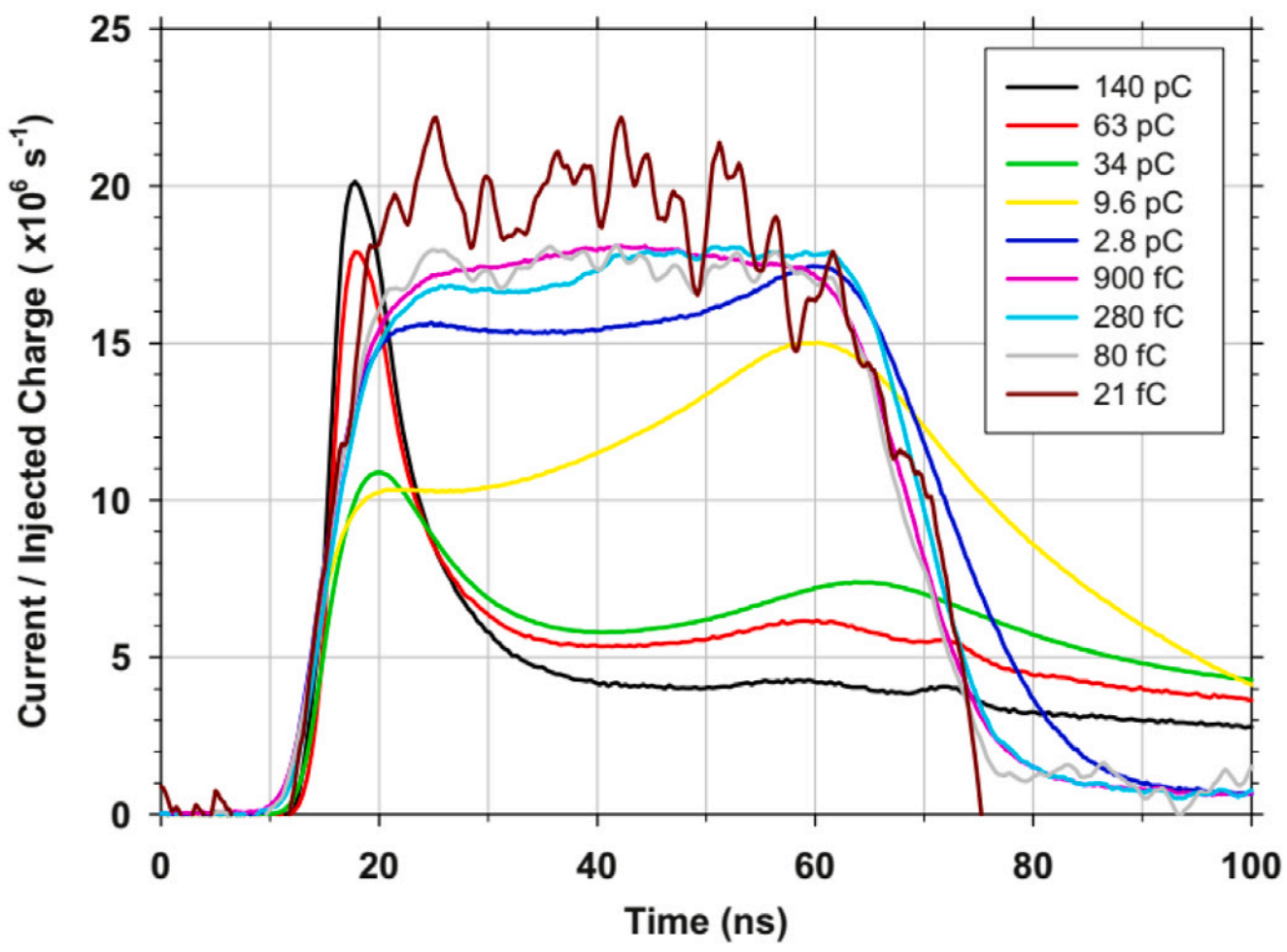

**Fig. 2.** Transient-current signals (holes) at 34 V for varying laser intensities, illustrating the transition from SCFC to SCLC currents at $Q_{inj} = 34$ pC.
*Source:* Figure adapted from [101].

*4.2.1. Space-charge-free and space-charge-limited transport*

In TCT measurements, a sharp cusp or kink in the transient-current signal is the hallmark of space-charge-free current (SCFC), reflecting a uniform internal electric field and minimal interaction among injected carriers. In contrast, the absence of a cusp, or the appearance of broadened or asymmetric transients, typically indicates the onset of space-charge-limited current (SCLC). Fig. 2 (adapted from [101]. see also [85] where space charge is impurity and polarization dependent) shows TCT signals (holes) at 34 V for varying laser intensities and illustrates the transition from SCFC to SCLC at $Q_{inj} = 34$ pC, beyond which the cusp is no longer observed.

Quantitatively, SCFC holds when the injected charge satisfies $Q_{inj} < CV_{bias}$, with C the sensor capacitance and $V_{bias}$ the applied bias voltage. Once $Q_{inj}$ approaches or exceeds $CV_{bias}$, SCLC sets in, the transient therefore broadens and loses its cusp [101].

SCLC can originate from several mechanisms, including the build-up of electron–hole plasma [101,102], excitons or electron–hole droplets [103], and carrier trapping due to defects or impurities [85,91]. These effects perturb the internal electric field, cause momentary or long-term polarization, and limit carrier mobility.

Transient time extraction in SCLC transient-current signals remains a challenging task, particularly in the absence of a well-defined cusp that clearly marks charge collection. Several theoretical frameworks have addressed this problem [104–106]. Isberg et al. [97] adopted the concept of an effective thickness and used the modified (sample to sample) approximations of $t_{eff} \approx 0.78 t_t$ from [106] to extract mobilities. [61,107] reported a slightly higher factor of $\approx 0.85\, t_t$. Further investigations are needed to reliably determine $t_{eff}$, as it directly impacts the extracted values of $v_d$ and $\mu_{eff}$.

*4.2.2. Polarization and field distortion*

Lattice defects, grain boundaries, and residual impurities introduce deep-level traps in diamond. These traps lead to persistent carrier trapping [82,91,108,109]. Trapped charge accumulates over time to form localized space-charge regions that progressively distort the applied electric field—a phenomenon known as polarization. During a TCT experiment, the sustained build-up of space charge increases the electric-field distortion, which ultimately degrades CCE, reduces signal fidelity, and, in severe cases, can lead to a complete loss of detector response [91]. Various mitigation strategies have been proposed to address polarization, including bias-polarity reversal, white-light illumination, and thermal annealing, all aimed at promoting trap depopulation and restoring electric-field uniformity [91,110–113].

**Table 2**
Room-temperature Hall mobilities of electrons and holes in diamond.

| Ref. | $l_x$, $l_y$ cm | $\mu_{H,e}/\mu_{H,h}$ ($cm^2$/Vs) | B (T) | $V_B$,$V_H$ (V), (mV) |
|---|---|---|---|---|
| [63] | 0.58, 0.35 | **900 ± 50/–** | 0.278 | 370, 360 |
| [64][a] | | **1800 ± 540/–** | 1.0 | 90–400, – |
| [64][b] | | –/>1200 | 1.0 | 90–400, – |
| [65] | 0.299, 0.153 | –/1550 ± 150 | 0.213 | – |
| [66] | 0.40, 0.20 | –/1140–1450 | 0.27–0.3 | – |
| [67] | | –/1550 ± 50 | 1 | – |
| [67] | | –/800 ± 80 | 1 | – |
| [68][c] | 0.6, 0.15 | **900–2000/–** | 3 | |
| [68][d] | 0.6, 0.15 | –/1500 | 3 | |

*Note:* Numbers preceding the slash (in bold) denote electron mobilities, numbers following the slash denote hole mobilities. The same convention (with comma) applies in the $V_B$, $V_H$ column.
[a] Seven samples.
[b] Seven samples.
[c] Eight *n*-type samples.
[d] One *p*-type sample.

*4.3. Interpreting transport curves:* $v_d(E)$ *and* $\mu_{eff}(E)$

Fig. 3(a) sketches a typical drift-velocity response $v_d(E)$ for diamond. Three regimes are evident, i) a low-field linear region, where API dominates and $v_d \propto E$, ii) a nonlinear transition around the critical field $E_c$, where API and OPI act concurrently, and (iii) a high-field saturation plateau limited primarily by OPI, in which $v_d$ saturates and is denoted by $v_{sat}$. The value of $v_{sat}$ is material specific. For intrinsic semiconductors such as silicon and diamond, the ultimate saturation of drift velocity has not been observed experimentally [60,114,115].

For electrons, the $v_{d,e}$, displays a noticeably broader saturation plateau than for holes, suggesting an additional scattering mechanism operating alongside OPI. The vertical dashed line marks $E_c$, the boundary between the linear and saturation regimes.

To emphasize that the same data convey mobility information, each $v_d(E)$ curve is divided by E and replotted as $\mu_{eff}(E)$ in Fig. 3(b), according to Eq. (21). In this interpretation, $E < E_c$ defines low electric field ($< 0.6$ V μm$^{-1}$) where $v_d$ scales linearly with $E$ (see Fig. 3(a)) resulting in constant low field mobility $\mu_0$ (see Fig. 3(b)) that is a material-specific constant and serves as a fundamental transport parameter characterizing the semiconductor material. The electric field range, $E > E_c$ defines the onset of saturation velocity $v_s$ (see Fig. 3(a)).

Throughout the detector operating (DO) range (defined as 0.1–4.0 V μm$^{-1}$ in this study), the electron drift velocity remains lower than the hole drift velocity, $v_{d,e} < v_{d,h}$, consistent with the mobility hierarchy $\mu_{0,e} < \mu_{0,h}$ established earlier.

*4.4. Measured mobilities and saturation velocities*

Historically, charge-carrier transport was first investigated using the Hall-effect measurement technique (see Table 2) in natural type IIa diamond. Redfield (1954) [64] investigated seven *p*-type samples. Dean (1965) [67] measured Hall mobility in three samples, one of which had also been studied by Austin (1956) [65]. One sample showed a very low mobility. Konorova (1967) [68] measured Hall mobilities for eight *n*-type and one *p*-type sample as functions of temperature and electric field, finding considerable variation among the *n*-type mobilities. Hall mobilities were extracted using Eq. (20) or a modified form thereof.

For the reasons discussed in Section 4.1, the diamond community moved to TCT-based experimental techniques to measure and characterize charge-carrier drift as a function of electric field and temperature in diamond. This section reviews the $\mu_0$ and $v_s$ values (Table 3) calculated or extrapolated using models consistent with TCT data analysis.

Pearlstein and Sutton [81] performed an $\alpha$-source charge-collection measurement in the Hecht/CCD framework and combined it with a CSA-based oscilloscope rise-time analysis to obtain electron and hole mobilities. The Hecht formula was used to extract the $\mu\tau$ product (termed "range" in their work) as

$$\frac{Q}{Q_0} \approx \frac{\mu\tau}{d^2} V \quad \Longrightarrow \quad \mu\tau \approx \frac{Q/Q_0}{V/d^2}, \tag{22}$$

where $Q/Q_0$ is the collected/deposited charge ratio, $V$ is the bias voltage, and $d$ is the detector thickness. The lifetime $\tau$ was obtained from the rising edge of the preamplifier (CSA) voltage pulse, and the mobility then follows from $\mu = (\mu\tau)/\tau$.

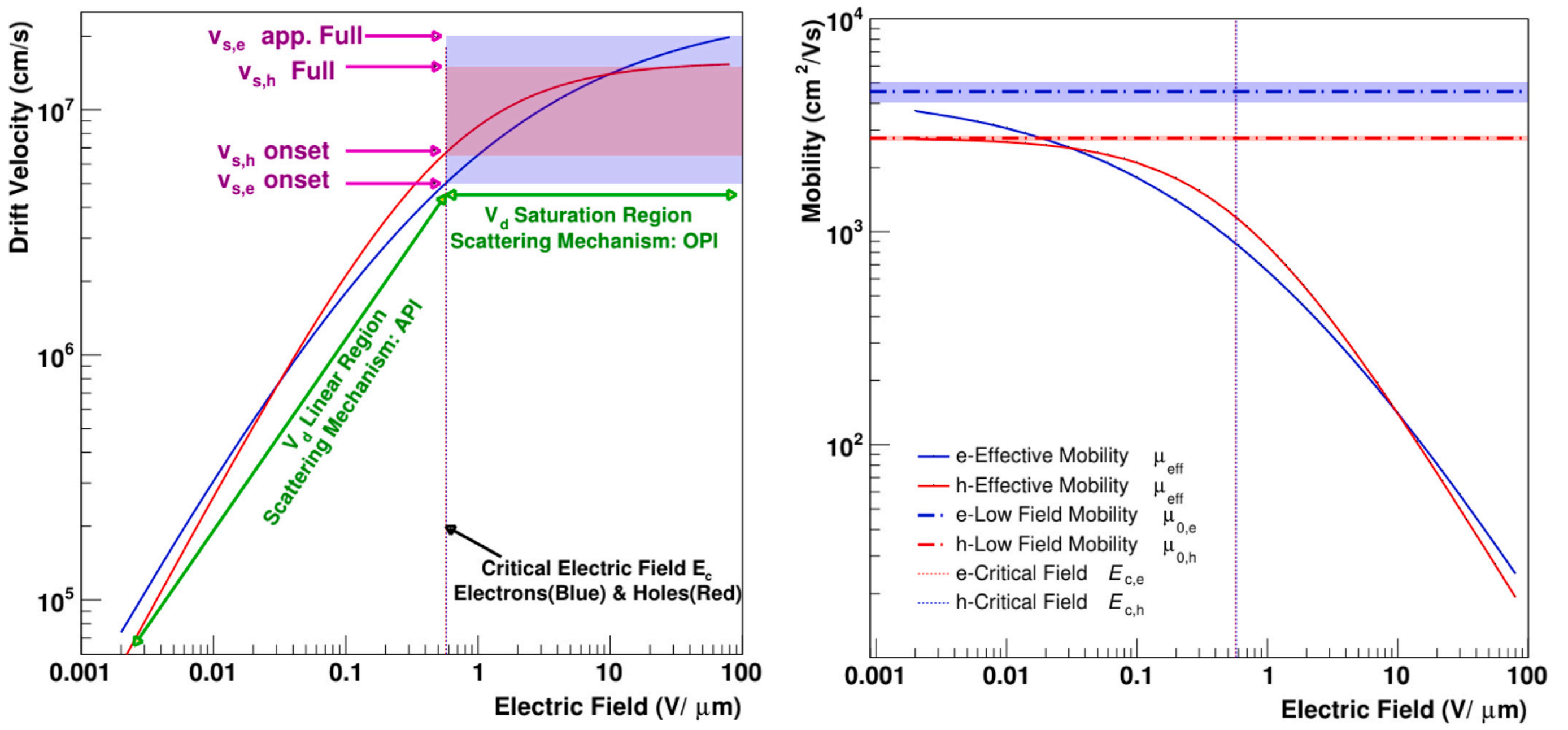

(a) (b)

**Fig. 3.** Drift velocity $v_d$ and effective mobility $\mu_{\mathrm{eff}}$ versus electric field E (schematic). (a) $v_d$–$E$ characteristics showing a linear (left) regime and a velocity-saturation (right, transparent) regime separated by the critical field $E_c$. The onset and termination of the saturation plateau bound the range of saturation velocities $v_s$ for electrons (blue) and holes (red). (b) $\mu_{\mathrm{eff}}$–$E$ characteristics illustrating the decrease of $\mu_{\mathrm{eff}}$ with increasing field. $\mu_{\mathrm{eff}}$ remains approximately constant up to $E = E_c$, after which it declines as velocity saturation sets in. Horizontal lines indicate the low-field mobilities $\mu_{0,e}$ and $\mu_{0,h}$. It must be noted that our analysis suggests a modified curve for electons (blue) where $v_{sat,e}, \mu_{0,e} < v_{sat,h}, \mu_{0,h}$, details are provided in the main text.

Nava & Canali (1979) [31] and Canali et al. (1979) [32] performed TCT experiments using a pulsed electron accelerator as the ionizing source (Table 3).

Reggiani et al. (1981) [34] characterized hole mobilities in eight samples of natural diamond, three along the [100] and five along [110] directions with thicknesses ranging from 60–420 μm, an area of 10 mm$^2$, and mean free drift times 7–17 ns. The experimental setup was the same as in [31,32].

Pan et al. (1992) [116] applied transient photoconductivity (laser-TCT) to type IIa natural diamond and pcCVD to extract mobilities and lifetimes. For natural diamond the mobility was calculated using

$$\mu = \frac{\mu\tau}{\tau}, \tag{23}$$

where $\mu\tau$ was proportional to the integrated charge, and $\tau$ was measured from the decay using $i(t) = i_0 e^{-t/\tau}$, with $i(t) \propto n(t)$, the carrier concentration. For pcCVD, electron–hole scattering was also considered and Matthiessen's rule was used,

$$\frac{1}{\mu} = \frac{1}{\mu_0} + \frac{1}{\mu_{cc}}. \tag{24}$$

$\mu_{cc}$ is given in Eq. (19). Note that the pcCVD films under test were very thin (6 and 3 μm), and electron and hole mobilities were not distinguishable in this case.

In a separate study, Pan et al. (1993) [94] characterized single-crystal natural type IIa diamond using (i) charged-particle-induced conductivity (PIC, CCE-type) in the bulk and (ii) time-resolved transient photoconductivity (TPC, TCT-type) near the surface. They concluded that surface and bulk transport mechanisms are the same. In PIC, both electrons and holes contributed to the signal and the ionizing particle traversed the crystal through a contact, i.e., the excitation was *parallel* to the field, whereas in TPC synchrotron X-rays impinged on the side surface, i.e., the excitation was perpendicular to the field. For the CCE-type PIC experiment,

$$S = \frac{Q_{col}}{E_{dep}} = \left(\frac{q}{E_{eh}}\right)\frac{\mu\tau E}{d} = \left(\frac{q}{E_{eh}}\right)\frac{CCD}{d}, \tag{25}$$

$$\Rightarrow CCD = \mu\tau E \Rightarrow \mu = \frac{CCD}{\tau}, \tag{26}$$

with $S$ the sensitivity and $E_{eh}$ the energy required to create an e–h pair. $\tau$ is independently extracted from the falling edge of the signal. Mobility extraction from the TCT-type TPC measurement follows Eq. (23).

Isberg et al. (2002) [97] carried out low-field TCT under SCLC conditions using a laser as the carrier-generation source (see Section 4.2.1) and analyzed the transients with a modified Drude model,

$$v_d = \mu E/\beta \quad \Longrightarrow \quad \mu = (V/\beta d^2)/t, \tag{27}$$

to extract low-field mobilities. Here, $\beta$ [106] is a correction factor that compensates for space-charge effects and device imperfections in diamond (see Section 4.2.1). Hole mobilities were independently verified using a *p–i* junction diode via the Mott–Gurney relation

$$\vec{J} = \frac{9}{8}\frac{V^2}{d^3}\epsilon\mu. \tag{28}$$

The temperature dependence of the hole mobility was also assessed from 300 to 540 K, and the carrier lifetime was discussed qualitatively.

Pernegger et al. (2005) [85] conducted TCT experiments to assess mobilities in scCVD diamond using an Am-241 $\alpha$ source as the ionizing radiation, ensuring operation in the SCFC regime (see Section 4.2.1). Data were analyzed with the TK model.

In addition, a modified Drude model,

$$v_d = 2\,\mu\,E\,\ln\!\left(\frac{V \pm V_c}{V \mp V_c}\right) \Longrightarrow \mu = \frac{d^2}{2\,t\,V_c}\,\ln\!\left(\frac{V \pm V_c}{V \mp V_c}\right), \tag{29}$$

was applied to directly extract $\mu_{\mathrm{eff}}$ and thus the low-field mobility. The factor $2\ln\big((\mathrm{V} \pm \mathrm{V_c})/(\mathrm{V} \mp \mathrm{V_c})\big)$ in Eq. (29) compensates for space charge and is comparable to the $\beta$ factor used in [61,97,106]. The measured values $\mu_{\mathrm{eff,e}} = 2500\,\mathrm{cm^2\,V^{-1}\,s^{-1}}$ and $\mu_{\mathrm{eff,h}} = 3000\,\mathrm{cm^2\,V^{-1}\,s^{-1}}$ at an applied electric field of $0.2\,\mathrm{V\,\mu m^{-1}}$ exceeded the low-field mobilities extracted using the TK model (Eq. (4)), $\mu_{\mathrm{o,e}} = 1714\,\mathrm{cm^2\,V^{-1}\,s^{-1}}$ and $\mu_{\mathrm{o,h}} = 2064\,\mathrm{cm^2\,V^{-1}\,s^{-1}}$ (see Table 3), and are consistent with Ref. [97]. Transient pulses without a sharp falling edge were excluded from the analysis. The bias voltage at which the cusp disappears is referred to as the compensation voltage ($V_c$) for space charge. This analysis also quantifies carrier lifetime and net effective space charge.

Tranchant et al. (2007) [117] reported similar mobilities using both laser and Am-241 sources, indicating that the laser intensity was sufficiently low to avoid the SCLC domain.

Bogdan et al. (2007) [107] assessed three samples (one IIa natural and two scCVD) with H and O terminations and Al contacts using laser-TCT, and directly extracted mobility at $0.22\,\mathrm{V\,\mu m^{-1}}$ via Eq. (27) with a slightly higher (than used by [97]) $\beta = 0.85$ (see Section 4.2.1). One CVD sample showed significantly lower electron and hole mobilities, attributed to additional scattering. The TK model (Eq. (4)) was used to extrapolate $\mu_{0,\mathrm{e/h}}$.

Nesládek et al. (2008) [61] extended the work of [107,117] using the same technique. They compared samples with different O, H terminations and Al, Au contacts, and investigated the transition from SCFC to SCLC. Hole mobility of laser-generated carriers was found to be lower than for $\alpha$-generated carriers. The lifetime was estimated to be 11 ns.

Pomorski (2008) [86] observed higher saturation velocities and mobilities after evaluating more than 30 scCVD samples using lower-energy $\alpha$'s (4.724 MeV) from an Am-241 source in a TCT setup. It was shown that $\mathrm{v_{s,e}} < \mathrm{v_{s,h}}$ in the field range $0.01\,\mathrm{V\,\mu m^{-1}} < E < 10\,\mathrm{V\,\mu m^{-1}}$ (see Fig. 3). Data were analyzed using the CT model (Eq. (5)). This work continues [118,119,122], which evaluated CVD samples using TCT. In 2005 [118] they reported $\tau_e = 174 \pm 15\,\mathrm{ns}$ and $\tau_h = 968 \pm 230\,\mathrm{ns}$. In 2006 [119], 15 samples showed $\tau_e = 165 - 321\,\mathrm{ns} \pm 20\,\%$ and $\tau_h = 150 - 968\,\mathrm{ns} \pm 10\,\%$.

Gabrysch et al. (2008) [95] used ultrashort hard X-ray pulses from the Sub-Picosecond Pulse Source (SPPS) at SLAC as the ionizing source to extract the transit time in TCT at very low fields. Mobilities were extracted using the Drude model (Eq. (2)). Gabrysch et al. (2011) [99] conducted laser-TCT experiments between 83 and 460 K and directly measured mobilities at very low fields ($0.009\,\mathrm{V\,\mu m^{-1}}$), the lowest room-temperature field in our review.

Jansen et al. (2012–2014) [62,120,123] used an alpha-TCT setup to evaluate low-field mobilities over a range of fields and temperatures. Sharp falling edges were observed at all voltages, indicating the absence of space charge. The TK model (Eq. (4)) was used to extract the parameters $\mu_0$ and $\mathrm{v_s}$, and the Drude model (Eq. (2)) was used to estimate the instantaneous $\mathrm{v_d}$ at $0.8\,\mathrm{V\,\mu m^{-1}}$.

Majdi et al. (2016) [121] performed TCT measurements to study conduction-band transport of valley-polarized electrons (see Section 6). Using Eq. (6) and assuming equipopulated valleys, they extrapolated the mobility to room temperature.

Kassel (2017) [88] characterized five scCVD samples using an alpha-TCT setup. The spread in $\mathrm{v_{d,h}}$ exceeded that in $\mathrm{v_{d,e}}$, contrary to the findings of Ref. [86]. One irradiated sample ($\Phi = 30.1 \times 10^{13}\, n_{1\,\mathrm{MeV}\,eq}\,\mathrm{cm^{-2}}$) showed the same $\mathrm{v_d}$ for electrons and holes as the unirradiated samples.

Berdermann et al. (2019) [89] investigated scCVD diamond-on-iridium (DOI) sensors with various contact types and configurations using an alpha-TCT setup. $\mathrm{v_s}$ and $\mu_0$ were extracted using the TK model. Significant spread and non-uniformity in charge-carrier drift were observed across the samples.

Bassi et al. (2021) [90] characterized 28 detectors for dosimetry and beam-loss monitoring at SuperKEKB using an alpha-TCT setup. The TK model (Eq. (4)) was used to extract $\mu_0$ and $\mathrm{v_s}$. Marked sample-to-sample differences in the TCT response—and in the derived parameters—were observed, attributable to crystal imperfections.

Zyablyuk et al. (2022) [91] studied polarization/ depolarization dynamics in scCVD diamond detectors and assessed polarized and unpolarized samples using an alpha-TCT setup. Data were analyzed using the TK model (Eq. (4)). Their results confirmed that scCVD can remain polarized for long periods in the absence of an external stimulus. A 365 nm, 1 W LED fully depolarized a strongly polarized detector in less than 300 ms.

Portier et al. (2023) [93] characterized scCVD diamond using an electron-beam–induced TCT experiment over 13–300 K (see Section 6). The TK model (Eq. (4)) was used for parameterization.

Kholili et al. (2024) [92] characterized an scCVD sample using alpha-TCT and analyzed the data with the TK model (Eq. (4)). Electron and hole lifetimes were reported as 169 ns and 490 ns, respectively. A TCAD device simulation for scCVD diamond was also implemented.

The brief review in this section shows that reported TCT mobilities in diamond do not exhibit a universal electron–hole hierarchy—both $\mu_{0,h} > \mu_{0,e}$ and $\mu_{0,e} \gtrsim \mu_{0,h}$ are observed. The spread in extracted $\mu_0$ and $v_s$ values arises primarily from (i) crystal type and material quality (deep traps and polarization), (ii) surface termination and contact metallurgy, (iii) transport regime and analysis assumptions (SCFC vs. SCLC; transit-time definitions), (iv) the electric-field window covered, (v) the ionizing source used ($\alpha$, laser, electron), and (vi) the choice of parameterization for $v_{\mathrm{d}}(E)$ (Drude/TK/CT models or modified variants).

## 5. Dependence of mobility on electric field

To mitigate the inter-study heterogeneity of published TCT drift-velocity data, this section compiles digitized $v_{\mathrm{d}}(E)$ curves from the literature and benchmarks compact transport parameterizations on a common dataset. Three models are evaluated (TK, CT, and PW) using (i) experiment-wise fits (Group 1) and (ii) pooled fits (Group 2) to obtain reference parameter sets for electrons and holes. Source-dependent normalization effects are treated by analyzing $\alpha$- and laser-TCT data separately and via a simple multiplicative cross-normalization. Based on the combined results, we adopt best-performing reference models and report their parameter values for use in simulation and comparison studies.

### 5.1. Models under consideration

We compare three candidate models, TK, CT, and a piecewise (PW) model (Eq. (30)) that we propose. The TK model is commonly used in TCT characterizations of diamond, whereas the CT model is widely used for silicon. We propose a piecewise model that captures the broad range of reported data and can serve as a standard within simulation frameworks.

$$v_d(E) = \begin{cases} \alpha\,E, & E < E_t, \\ \dfrac{v_s\,E}{E_c\,\left(1 + (E/E_c)^{\beta}\right)^{1/\beta}}, & E \geq E_t. \end{cases} \tag{30}$$

Here $\alpha \equiv \mu_{0,1}$ denotes the low-field mobility. Electron drift over a wide range of electric fields is well described by a linear regime followed by a hill-type non-linear region that ultimately saturates (see Fig. 3). The linear regime extends up to the threshold field $E_t$, fixed at $0.08\,\mathrm{V\,\mu m^{-1}}$ for electrons and $0.065\,\mathrm{V\,\mu m^{-1}}$ for holes throughout this work (shown as magenta vertical line in Figs. 7 (a,b,c,d)), and is parameterized by the Drude Model (Eq. (2)). The interval $E > E_t$ defines the hill-type non-linear region that can be described by the CT model. PW model is the approximation of more general and fundamental electron transport framework discussed in Section 6.

**Table 3**
Reported low-field mobilities and saturation velocities for electrons and holes in diamond.

| Ref. | # | E range (V µm$^{-1}$) | | Model | $\mu_0$ (cm$^2$/Vs) | | $v_s$ ($10^6$ cm s$^{-1}$) | |
|---|---|---|---|---|---|---|---|---|
| | | Electrons | Holes | | $\mu_{0,e}$ | $\mu_{0,h}$ | $v_{s,e}$ | $v_{s,h}$ |
| [81] | 2 | 0.250 – ——— | 0.250 – ——— | (22) | 3900 ± 585 | 4800 ± 960 | —— | – |
| [116][1] | 1 | ——— – ——— | ——— – ——— | (23) | 500–3000 | —— | —— | – |
| [116][2] | 1 | ——— – ——— | ——— – ——— | (24) | ~43,50 | —— | —— | – |
| [94] | 1 | ——— – ——— | ——— – ——— | (26) | 1500 | 1000 | —— | – |
| [31] | – | 0.052 – 5.400 | 0.060 – 3.96 | – | 2400 | 2100 | 15 | 10.5 |
| [32] | – | 0.052 – 5.400 | 0.060 – 3.96 | – | 2400 | 2100 | 15 | 10.5 |
| [34] | 3 | ——— – ——— | 0.073 – 4.20 | – | —— | 2080 – 2140 | – | 11 ± 0.1 |
| [34] | 5 | ——— – ——— | 0.067 – 5.21 | – | —— | 2050 – 2150 | – | 11 ± 0.1 |
| [97] | 4 | 0.018 – 0.400 | 0.018 – 0.40 | (27) | 4500 | 3800 | – | – |
| [97] | 4 | ——— – ——— | ——— – ——— | (28) | – | 4090 | – | – |
| [98] | >1 | ——— – ——— | ——— – ——— | (2) | – | 3400 ± 400 | – | – |
| [85] | 1 | 0.220 – 1.480 | 0.200 – 0.80 | (4) | 1714 | 2064 | 09.60 | 14.1 |
| [85][3] | 1 | 0.220 – 0.800 | 0.200 – 0.80 | (29) | 2500 | 3000 | 05.60 | 07.5 |
| [118] | 1 | 0.168 – 0.848 | 0.068 – 1.36 | (4) | 2071 ± 212 | 2630 ± 123 | 8.5 ± 0.8 | 13.4 ± 0.5 |
| [119] | 15 | 1.580 – 07.71 | 0.867 – 7.92 | (4) | 1300–3100* | ≈2330 ± 250 | ≈19 | ≈14 |
| [117] | 1 | ——— – ——— | 0.060 – 0.78 | – | 2500 ± 50 | 2050 ± 50 | – | – |
| [107][3] | 2 | 0.220 – ——— | 0.220 – ——— | (27) | ~2650 | ~2150 | – | – |
| [107][3] | 1 | 0.220 – ——— | 0.220 – ——— | (27) | 2050 | 2250 | 10 | 12 |
| [107] | 1 | 0.120 – 0.225 | 0.120 – 0.22 | (4) | ~950 | ~650 | – | – |
| [61] | 1 | ——— – ——— | 0.120 – 0.22 | (4) | 2050 | 2250 | 12 | 10 |
| [61][4] | 1 | 0.023 – 1.140 | 0.045 – 1.14 | (4) | ~2500 ± 300 | ~2250 ± 300 | – | – |
| [86] | >30 | 0.101 – 11.20 | 0.031 – 11.2 | (5) | 4551 ± 500 | 2750 ± 70 | 26.3 ± 2.0 | 15.7 ± 1.4 |
| [86][3,5] | >30 | 0.100 – 09.00 | 0.060 – 11.0 | (2) | 1800 | 2450 | 14.3 | 14.3 |
| [95] | 2 | 0.029 – 0.654 | 0.058 – 0.15 | (2) | ≥2760 | ≥2750 | – | – |
| [99] | 3 | 0.009 – 0.416 | 0.009 – 0.40 | (4) | 2296 ± 39 | 2322 ± 43.6 | 07.12 ± 0.074 | 10.8 ± 0.12 |
| [120] | 1 | 0.077 – 1.700 | 0.076 – 1.70 | (4) | 1440 ± 220 | 2280 ± 110 | 09.9 ± 1.1 | 12.5 ± 1.1 |
| [120][6] | 1 | ——— – 0.800 | ——— – 0.80 | (2) | – | —— | 05.21 ± 0.22 | 07.39 ± 0.13 |
| [62] | 3 | 0.077 – 1.700 | 0.076 – 1.70 | (4) | 1850 ± 23 | 2534 ± 20 | 13.2 ± 0.3 | 14.2 ± 0.2 |
| [62] | 3 | ——— – 0.940 | ——— – 0.94 | (2) | – | —— | 06.28 ± 0.04 | 08.69 ± 0.05 |
| [87] | 1 | 0.226 – 0.676 | 0.112 – .676 | (5) | – | —— | – | – |
| [109] | 1 | ——— – ——— | 0.100 – 0.80 | (4) | – | 2972 | – | 12.3 |
| [121] | 1 | .0003 – 0.069 | ——— – ——— | ——+ | 2540 | – | – | – |
| [121] | 1 | ——— – ——— | ——— – ——— | (21) | 2200 | – | – | – |
| [88] | 5 | 0.176 – 2.300 | 0.176 – 2.30 | (5) | 14948 ± 8303 | 2615 ± 148 | 64.7 ± 37.5 | 15.3 ± 10.8 |
| [88][7] | irr.1 | 0.220 – 1.820 | 0.475 – 1.82 | Inf. | 14948 | 2615 | 64.6 | 15.3 |
| [89] | 7 | 0.096 – 4.000 | 0.033 – 4.00 | (4) | 1150 – 2276 | 1750 – 3080 | 10.0 – 15.0 | 14.0 – 17.0 |
| [100] | 1 | 0.555 – 2.300 | 0.364 – 1.38 | (4) | 1300 | 2000 | 11.0 | 13.0 |
| [100][8] | 1 | 0.555 – 2.300 | 0.364 – 1.38 | (5) | 5130 ± 995 | 2270 ± 216 | 26.0 ± 4.08 | 14.3 ± 0.96 |
| [90] | 28 | 0.300 – 1.600 | 0.300 – 1.60 | (4) | 1700 | 2000 | 09.00 | 13.0 |
| [91] | 2 | 0.164 – 1.090 | 0.164 – 1.09 | (4) | 2022 ± 156 | 2238 ± 121 | 08.05 ± 0.505 | 12.06 ± 0.70 |
| [93] | 1 | 0.028 – 0.916 | 0.018 – 0.91 | (4) | 1610 ± 10 | 2190 ± 30 | – | – |
| [92] | 1 | 0.084 – 0.820 | 0.082 – 0.80 | (4) | 2205 | 2559 | 07.60 | 13.5 |

*Note:* # denotes the number of samples. [116][1] Mobility is $\mu_{0,e+h}$ (sum of electron and hole mobilities). [116][2] pcCVD samples where e–h scattering is dominant. [85,86,107][3] Directly measured mobility values at the lowest electric field. [61][4] Various termination–contact combinations, see main text. [86][5] Initial-field mobility is directly measured, the final field corresponds to the instantaneous velocity at that point. [120][6] Instantaneous $v_d$ at the given field. [88]7 "irr. 1" denotes an irradiated sample, "Inf." denotes inferred. [100][8] CT fitted by us, see main text. [121]* TK model (Eq. (4)) used at low T and extrapolated to room temperature via Eq. (6) to obtain the low-field mobility. [119]+ 1300–3100 ± 600.

### 5.2. Literature data and analysis strategy

We digitized literature data from 17 papers (for $v_d - E$) using PlotDigitizer PRO (https://plotdigitizer.com), with a digitization uncertainty of approximately ±1%. The digitized data, made available in the accompanying TUDOdata dataset [124], were organized into two distinct groups for analysis.

**Group 1** comprises data organized by individual experiments. Models were evaluated separately using the data from each experiment. Detailed results are provided in Table 4.

**Group 2** pools data from all experiments and is further divided into Electron and Hole subgroups. In these subgroups, models were tested over two electric-field ranges, (i) the full range (0.009–1.2 V µm$^{-1}$) and (ii) the detector-operating (DO) range (0.1–4.0 V µm$^{-1}$). Most TCT experiments fall within the DO range, which also encompasses the typical operational electric field used in HEP detectors (1.0 V µm$^{-1}$) (see Fig. 5). In Group 2, Within each electric-field range, data were grouped by source, (a) All (Alpha, laser, electron), (b) All$_{LA}$ (laser data scaled to alpha) (c) All$_{LA}$ (alpha data scaled to laser) (d) Alpha, and (e) Laser sources. There is only one TCT data set with electron source for electron and holes in scCVD diamond [93] and one TCT data set for holes in natural diamond [34].

Measurement uncertainties are either not reported or not digitizeable for most of the literature data, therefore, we adopt the digitization uncertainty (±1%) as the sole source of error in all cases.

Digitized data from sources [61,89,97] were excluded from the electron-group analysis (see Fig. 4), and data from sources [34,89,97] were excluded from the hole-group analysis. An initial evaluation identified data reported by Isberg et al. (2002) [97] as extreme outliers across all models for both electrons and holes. Data from Nesladek et al. (2008) [61] showed atypical electron-transport behavior at low electric fields due to different combinations of terminations and contact types, resulting in significant residual variation and outlying behavior across all models. Similarly, Berdermann et al. (2019) [89] used different substrates (DOI, see Section 4.4) and contact types, causing substantial variability in charge-carrier transport, reflected by an extremely uneven distribution of residuals, except for hole transport of DOI samples, which is included in the analysis. Reggiani et al. (1981) [34] data for hole transport in the [110] lattice direction are neglected because the reported $v_d(E)$ at higher electric fields differed significantly from hole transport in the [100] direction.

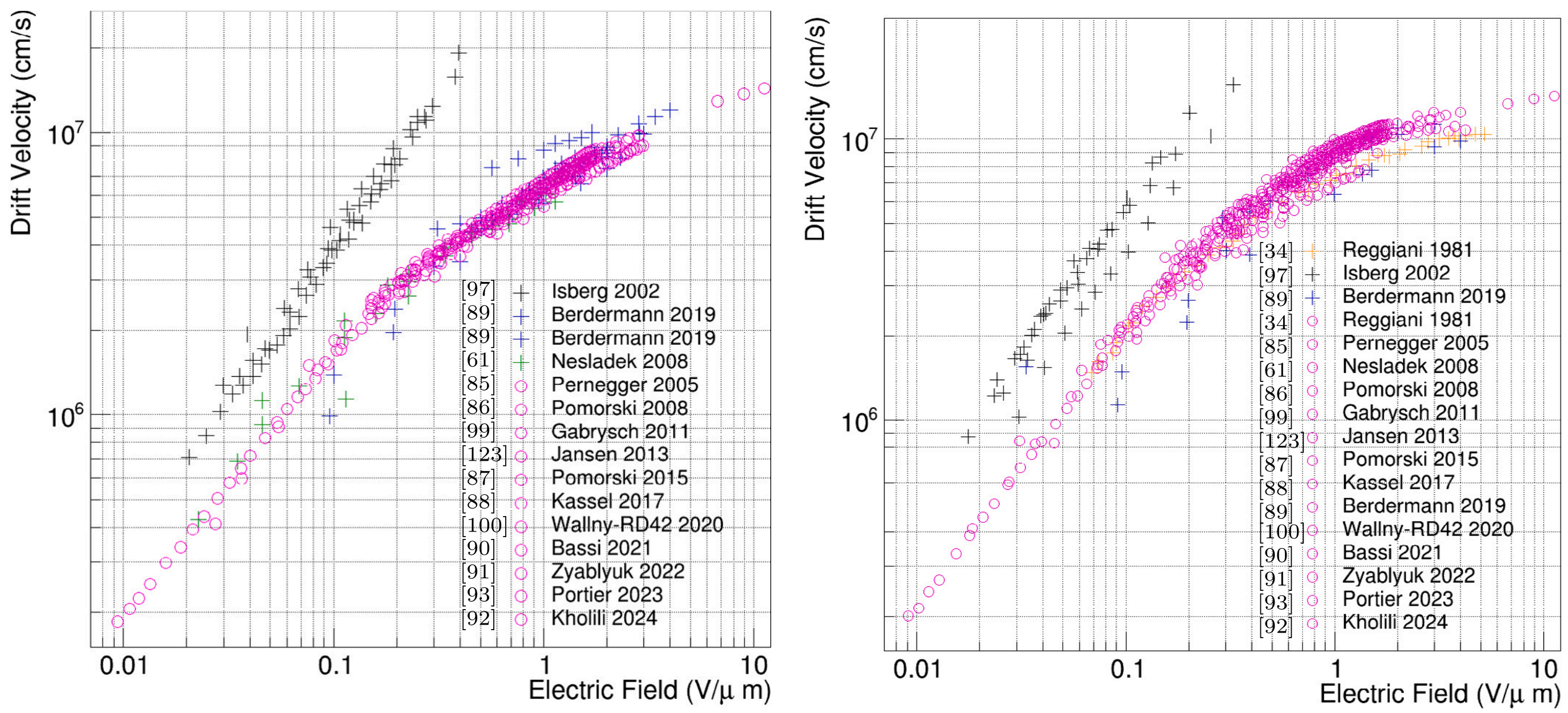

(a) (b)

**Fig. 4.** Selected (pink) and rejected data used for model evaluation of (a) electron and (b) hole transport.

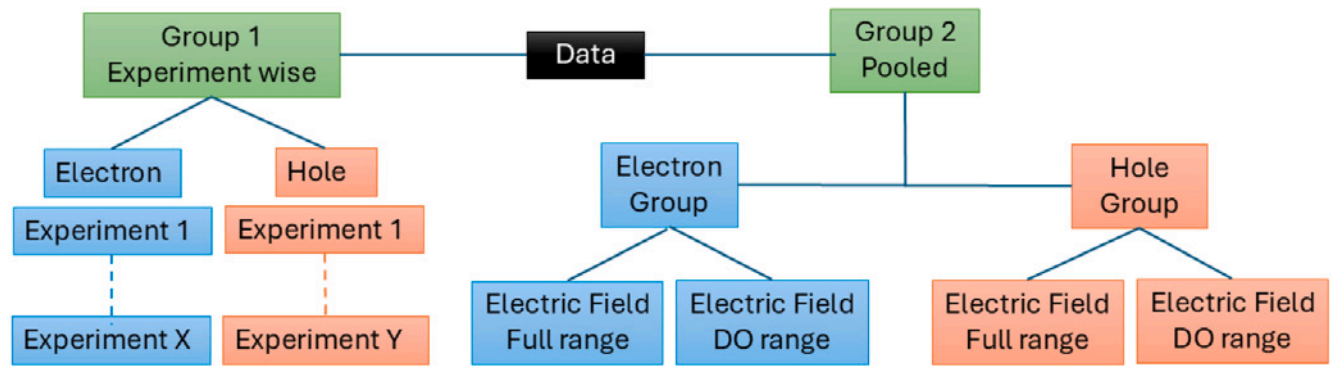

**Fig. 5.** Data grouping structure: Group 1 (experiment-wise) and Group 2 (pooled-data).

As the data are digitized from published figures where uncertainties were largely either not reported or not digitize-able (with digitization uncertainty being the only quantifiable error source), we evaluated candidate models using a suite of complementary metrics rather than relying on a single goodness-of-fit measure. We report reduced chi-square $\chi^2/\mathrm{ndf}$, coefficient of determination $R^2$ ($0 < R^2 < 1$), standardized residuals (z-score) $\sigma_z$, Akaike Information Criterion (AIC), and Bayesian Information Criterion (BIC). For an adequate fit, $\chi^2/\mathrm{ndf}$, $R^2$, and $\sigma_z \to 1$. For AIC and BIC, lower values indicate models that fit the data better. Pairwise comparisons use differences $\Delta$(A)BIC = (A)BIC$_x$ − (A)BIC$_y$, $\Delta$BIC of 6–10 indicates strong support for model y over x, and $\Delta$(A)BIC > 10 indicates very strong support. We adopted this multi-metric strategy because, in the absence of reported measurement uncertainties, any single metric can be misleading. Convergence of independent indicators ($\chi^2/\mathrm{ndf}$, $R^2$, and $\sigma_z \to 1$, and minimal AIC/BIC) provides a more robust basis for model selection. We report only $\chi^2/\mathrm{ndf}$ and BIC ($\Delta$BIC) in this section (Tables 4 and 5) due to space limitations.

In addition, to provide an intuitive description of typical model errors, we summarize the distribution of the relative residuals $r_i = (y_i^{\mathrm{mod}} - y_i^{\mathrm{data}})/y_i^{\mathrm{data}}$ over four electric-field ranges, global (0.009–12 V μm$^{-1}$), low-field ($E \leq 0.1$ V μm$^{-1}$), mid-field (0.1–4.0 V μm$^{-1}$), and high-field ($E \geq 4$ V μm$^{-1}$). For each range we look at (i) the median of the signed relative residuals, (ii) the central 68% and 95% intervals of the signed residuals, and (iii) the 68% and 95% quantiles of the absolute relative residuals, together with the global mean and standard deviation of the relative residuals.

### 5.3. Source effect on drift

Fig. 6(a,b) shows that, in most cases, $v_d$ measured with $\alpha$-TCT exceeds $v_d$ measured with laser-TCT. For electrons this holds across the full field range (Fig. 6a), for holes the difference is not consistent but clearly apparent below ~0.1 V μm$^{-1}$ and above ~0.3 V μm$^{-1}$V μm$^{-1}$.

A likely contributor to the systematic offset between $\alpha$-TCT and laser-TCT drift-velocity curves is the different charge-generation depth and injection density. Am-241 $\alpha$ particles deposit their energy within a shallow region of the diamond bulk (range ~10–15 μm), producing a spatially compact, locally high-density e–h packet. Such dense charge packets can transiently screen the applied field and enhance the role of ambipolar drift–diffusion and recombination, which modifies the early transient shape and can bias TCT–based drift extraction if a purely drift-dominated interpretation is assumed. In contrast, UV laser excitation (often 213–266 nm) generates carriers close to the illuminated surface, making the transient more sensitive to near-surface scattering/trapping and surface-termination-dependent defect states. Moreover, under near-band-edge optical excitation in diamond, a non-negligible fraction of the photoexcited population can reside transiently in excitonic states, with a measurable balance between free carriers and excitons, this can reduce the prompt free-carrier density relevant for drift. At high ionization rates, additional polarization/space-charge build-up can locally reduce the internal field and increase recombination, consistent with rate-dependent signal degradation reported for diamond sensors [88].

Fig. 6 also shows that the compiled literature provides limited overlap in field coverage between excitation methods. In reported dataset, $\alpha$-TCT points are absent below ~0.07 V μm$^{-1}$ (electrons) and ~0.03 V μm$^{-1}$ (holes). This is qualitatively consistent with the expectation that, at low fields, longer collection times increase the influence of diffusion and recombination, which can suppress the measurable prompt component and complicate $t_t$ extraction. Conversely, in the digitized literature we find no laser-TCT data above ~2.3 V μm$^{-1}$ (electrons) and ~1.6 V μm$^{-1}$ (holes).

The systematic $v_d$ offset between fits is evident in Fig. 6(a,b) for PW model fits. $\alpha$-TCT datasets in [85,90,92] lie close to the laser-TCT curve, the corresponding samples contain defects/impurities that build up space charge, which is consistent with our interpretation.

To calculate the offset between the drift-velocity datasets obtained with $\alpha$ sources and with laser excitation, we assumed that both probe the same underlying $v_d(E)$ curve up to an overall multiplicative normalization. Within each dataset we replaced all data points that share the same $E$ with a single entry, the arithmetic mean at that $E$. We then restricted both groups to their common field interval and linearly interpolated the laser drift velocities onto the electric-field grid of the pooled alpha dataset, yielding paired points $(v_{\mathrm{laser},i}, v_{\alpha,i})$.

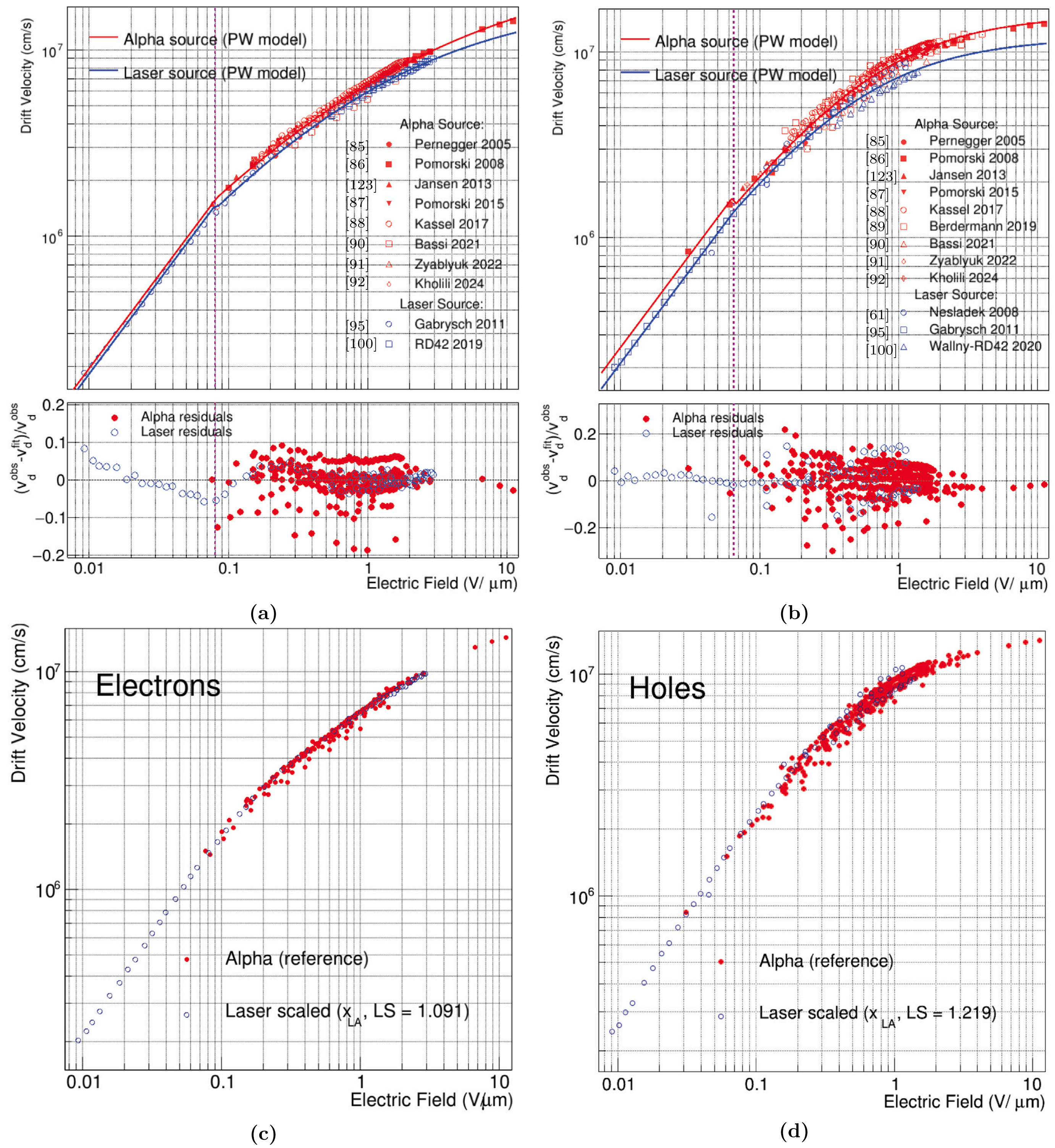

**Fig. 6.** $v_d - E$, grouped by excitation method ($\alpha$ and laser sources). (a,c) The $v_{d,e/h}$ measured with the $\alpha$-source (red) (Fitted with PW model just for illustration purpose) is systematically higher than that obtained with the laser. In $\alpha$-source measurements, diamond samples having defects show a reduction in $v_{d,e/h}$—similar to that observed under laser excitation—relative to defect-free material. (b,d) Fits using TK, CT and PW models describe both data sets, the $\alpha$-source $v_{d,e}$ exceeds the laser-derived $v_{d,e}$ by an approximately field-independent multiplicative factor.

Under the working hypothesis,

$$v_\alpha \simeq x_{\mathrm{LA}}\, v_{\mathrm{laser}},$$

the laser→ $\alpha$ scale factor $x_{\mathrm{LA}}$ was obtained from an unweighted least-squares fit,

$$x_{\mathrm{LA}} = \sum_i \left(x\, v_{\mathrm{laser},i} - v_{\alpha,i}\right)^2 = \frac{\sum_i v_{\alpha,i}\, v_{\mathrm{laser},i}}{\sum_i v_{\mathrm{laser},i}^2}.$$

The inverse mapping, treating the laser scale as reference,

$$v_{\mathrm{laser}} \simeq x_{\mathrm{AL}}\, v_\alpha,$$

was obtained analogously as

$$x_{\mathrm{AL}} = \frac{\sum_i v_{\mathrm{laser},i}\, v_{\alpha,i}}{\sum_i v_{\alpha,i}^2}.$$

As a robust cross-check, less sensitive to outliers and any residual field dependence, we also studied the pointwise ratios

$$r_{\mathrm{LA},i} = \frac{v_{\alpha,i}}{v_{\mathrm{laser},i}}, \qquad r_{\mathrm{AL},i} = \frac{v_{\mathrm{laser},i}}{v_{\alpha,i}},$$

and quote their median, mean, standard deviation, and maximum deviation from the median.

For electrons, using the common electric field interval 0.0767–2.9630 V µm$^{-1}$, the laser→ $\alpha$ least-squares factor is $x_{\mathrm{LA}} = 1.091$ (see Fig. 6(c)), while the pointwise ratios $v_\alpha/v_{\mathrm{laser}}$ have a median of 1.096 and a relative standard deviation of 4.6% (maximum deviation 16.5% from the median). The inverse fit (alpha→laser) yields $x_{\mathrm{AL}} = 0.915$, a median ratio $v_{\mathrm{laser}}/v_\alpha = 0.913$, and a relative spread of 5.0% with a maximum deviation of 19.8%. Thus, for electrons a single global normalization factor describes the overall offset between alpha and laser measurements at the few–percent level, but still leaves residual $E$–dependent structure of about (10–20%) level.

For holes, over the common electric field interval 0.0311-1.3827 V µm$^{-1}$, the laser→ alpha least-squares factor is $x_{\mathrm{LA}} = 1.219$ (see Fig. 6(d)). The ratios $v_\alpha/v_{\mathrm{laser}}$ have a median of 1.211 and a relative standard deviation of 9.8%, with a maximum deviation of 30.7% from the median. In the opposite direction the regression gives $x_{\mathrm{AL}} = 0.813$, a median ratio $v_{\mathrm{laser}}/v_\alpha = 0.826$, and a relative spread of 10.9% (maximum deviation 44.4%). These sizable ($\gtrsim 10\%$) variations indicate that, for holes, a single global scale factor provides only an approximate description of the relative normalization between alpha and laser datasets.

Fig. 6(c,d) shows the laser data scaled to alpha for electrons and holes. It is evident that the scaling quality is much better for electrons than holes. The relative more spread of laser data in holes is attributed to the lower $v_d$ values reported by Ref. [100].

### 5.4. Experiment wise (Group 1) model benchmarking

An experiment-wise analysis shows that the CT model describes the data best for most experiments. Table 4 reports model-performance metrics for charge-carrier drift-velocity data taken from 14 experiments. Although much of the charge-carrier transport TCT literature has been analyzed using the TK model, this analysis strongly supports the use of the CT model over short to moderate electric-field ranges for specific source types (see Table 3 for field ranges and Fig. 6 for source types).

It can be seen that data from Nesladek et al. [61] (electrons) and Berdermann et al. [89] performing worse for the reasons discussed in Section 5.2. Data of electron drift in Portier et al. [93] for the reason not known to us at least. They used electrons from Electron Microscope as an ionizing source generating e–h pairs in the diamond sample under test.

**Table 4**
Experiment-wise model-performance metrics for electron and hole drift. Here $\Delta\mathrm{BIC} = \mathrm{BIC}_{\mathrm{TK}} - \mathrm{BIC}_{\mathrm{CT}}$ is reported in place of individual BIC values due to space constraints. For details see Section 5.2.

| Exp. | Electrons | | | Holes | | |
|---|---|---|---|---|---|---|
| | $\chi^2$/ndf | | ΔBIC | $\chi^2$/ndf | | ΔBIC |
| | CT | TK | | CT | TK | |
| [34][a] 1981 | – | – | – | 8.36 | 11.51 | 149.5 |
| [34][b] 1981 | – | – | – | 6.70 | 12.37 | 235.4 |
| [85] 2005 | 0.50 | 9.75 | 81.4 | 0.55 | 2.98 | 22.4 |
| [61] 2008 | 164.29 | 169.17 | 321.7 | – | – | – |
| [86] 2008 | 4.80 | 71.71 | 3747.7 | 7.29 | 19.71 | 487.7 |
| [99] 2011 | 1.90 | 2.47 | 13.3 | 3.92 | 5.46 | 40.8 |
| [62] 2013 | 6.24 | 28.06 | 243.7 | 0.82 | 5.67 | 46.8 |
| [87] 2015 | 0.11 | 0.53 | −0.2 | 1.74 | 10.77 | 36.1 |
| [88] 2017 | 9.00 | 19.12 | 1794.6 | 18.99 | 19.37 | 95.5 |
| [89][c] 2019 | 163.07 | 168.12 | 311.3 | 134.01 | 137.65 | 279.4 |
| [89] 2019 | 118.73 | 114.44 | 38.5 | 441.02 | 427.92 | 162.7 |
| [100] 2019 | 2.10 | 4.75 | 125.3 | 3.43 | 3.55 | 2.5 |
| [90] 2021 | 1.38 | 14.22 | 50.9 | 0.13 | 1.49 | 3.8 |
| [91] 2022 | 1.56 | 6.03 | 52.5 | 3.14 | 4.25 | 13.8 |
| [93] 2023 | 31.34 | 53.66 | 252.1 | 5.10 | 4.70 | −2.3 |
| [92] 2024 | 2.23 | 2.15 | −1.6 | 1.38 | 3.53 | 26.6 |

Note:
[a] — [100] lattice direction.
[b] — [110] lattice direction.
[c] —Diamond on Iridium (DOI).

### 5.5. Pooled data (Group 2) model benchmarking

#### 5.5.1. Full field range

**Electrons:** Parameterization of electron mobility in diamond is particularly challenging because the reported drift-velocity data deviate from the usual linear-to-saturation pattern. The models are evaluated on a pooled dataset from 11 experiments, using digitized $v_d(E)$ values covering electric fields from 0.009 V µm$^{-1}$ up to 10.2 V µm$^{-1}$. The first block of Table 5 summarizes the performance metrics the models under consideration.

Differences between the models are evident in the $v_d$–$E$ characteristics shown in Fig. 7(a). For the “All sources” subset, all models reproduce the low- and mid-field behavior (0.1–4.0 V µm$^{-1}$) reasonably well. The median signed relative residuals in this range are at the level of 0.1–1.4%, and 68% of the absolute relative errors remain below about 6% for all three parameterizations. At high fields ($E \geq 4$ V µm$^{-1}$), however, the model differences become pronounced. The TK and CT models exhibit strong systematic overestimation, with median signed biases of 30.5% and 19.7%, respectively, whereas the PW model retains a much smaller high-field bias of 8.1%.

An analysis of the global (0.009-12 V µm$^{-1}$) residual distributions for the “All sources” subset further highlights these differences. For 68% of the measurements, the absolute relative errors are at the ∼6% level for all three models, but the PW model has a substantially narrower high-error tail, 95% of its absolute relative errors are below 9.5%, compared to 12.3% (CT model) and 13.4% (TK model). In other words, while the models behave similarly in the mid-field region, the PW parameterization provides a more accurate description of the data over the full field range, especially at high fields where the TK and CT models systematically overshoot the measurements.

These metrics provide strong support for the PW model in the “All” and “Laser” source subsets, whereas, for the “Alpha source” subset, the PW and CT models perform comparably. Overall, the PW model therefore offers a robust, generalized framework for describing electron drift over the full electric-field range.

The same analysis workflow was applied to these scaled subsets “All$_{\mathrm{LA}}$” and “All$_{\mathrm{AL}}$” and the corresponding performance metrics are reported in the first block (4th and 5th columns) of Table 5.

**Table 5**
Model performance for drift-velocity fits to TCT data, comparing the TK, CT, and PW models.

| Carrier E-Range | Model | All | | $All_{LA}$ | | $All_{AL}$ | | Alpha | | Laser | |
|---|---|---|---|---|---|---|---|---|---|---|---|
| | | $\chi^2$/ndf | BIC | $\chi^2$/ndf | BIC | $\chi^2$/ndf | BIC | $\chi^2$/ndf | BIC | $\chi^2$/ndf | BIC |
| Electrons | TK | 48.15 | 18 742 | 48.15 | 18 742 | 48.39 | 18 835 | 37.60 | 11 254 | 30.90 | 2357 |
| (0.009–12) | CT | 38.10 | 14 800 | 27.93 | 10 853 | 27.94 | 10 858 | 14.87 | 4447 | 10.75 | 820 |
| (V/μm) | **PW** | **29.38** | **11401** | **20.76** | **8065** | **20.81** | **8085** | **14.90** | **4455** | **6.30** | **488** |
| Holes | TK | 78.87 | 39 685 | 52.48 | 26 409 | 53.44 | 26 893 | 50.32 | 18 481 | 48.87 | 3527 |
| (0.009–12) | **CT** | **78.69** | **39523** | **50.43** | **25332** | **51.27** | **25755** | **48.00** | **17587** | **48.94** | **3488** |
| (V/μm) | PW | 76.33 | 38 274 | 50.56 | 25 360 | 51.42 | 25 792 | 48.00 | 17 550 | 48.60 | 3423 |

Fig. 7(b) shows the corresponding model fits for the laser-to-alpha scaled subset $All_{LA}$, the global median signed residuals are close to zero for all models (0.30% for TK, 0.12% for CT, and 0.42% for PW). Over the full field range 0.009–12 V μm$^{-1}$, 68% of the absolute relative errors are below 6.09% (TK), 4.30% (CT), and 3.41% (PW), indicating a clear tightening of the residual distributions compared to the unscaled "All sources" case. In the mid-field region (0.1–4.0 V μm$^{-1}$), which is most relevant for typical detector operation, the PW model again yields the best performance, with 68% of its absolute relative errors below 3.04%, compared to 4.11% for CT and 5.69% for TK. At high fields ($E \geq 4$ V μm$^{-1}$), the TK and CT models retain substantial positive biases (median signed errors of 29.52% and 13.22%, respectively), whereas the PW model is nearly unbiased (median 1.25%) and achieves very small 68% absolute relative-error bounds (1.68%, compared to 13.61% and 30.34% for CT and TK).

The Alpha-to-Laser scaled subset $All_{AL}$ produces very similar residual distributions and fit qualities. For example, the PW model exhibits a global median signed residual of 0.39%, a 68% absolute relative-error bound of 3.38% (3.07% in the mid-field region), and high-field 68% absolute errors below 1.57%. Nonetheless, across all three models the Laser-to-Alpha scaling systematically yields slightly lower $\chi^2$/ndf and information criteria. For the PW parameterization, $\chi^2$/ndf decreases to 20.76 with BIC = 8064 for "$All_{LA}$", compared to 20.81 and BIC = 8084 for "$All_{AL}$".

On this basis, we adopt the PW parameter set obtained from the Laser-to-Alpha scaled subset "$All_{LA}$" as our reference electron parameterization (see Table 11, 1st row). This choice is motivated by two considerations (i) it provides the best overall goodness-of-fit (lowest $\chi^2$/ndf, AIC and BIC), that exhibits minimal global and high-field biases together with the tightest mid-field error bounds, and (ii) it is anchored to the statistically more robust alpha-source data, which comprise 8 individual datasets (experiments) with more than 45 detector samples and 301 data points, compared to only two laser datasets with fewer than five diamond detectors and 90 data points. Parameter sets for other experimental source configurations can then be obtained by applying a simple multiplicative scaling factor $x$ to this reference parameterization. The extracted parameter values of $v_s$, $E_c$, $\mu_{0,1}$, and $\beta$ for all the three models are listed in Table 12 (see Appendix A).

**Holes:** Hole transport in diamond is more uniform than electron transport, owing primarily to phonon scattering and the absence or strong suppression of additional lattice-scattering mechanisms at room temperature. This leads to a typical linear-saturation pattern in the $v_d$–$E$ curves that can be described by Hill-type functions (TK and CT), as reflected in Table 5 (2nd block). The models were evaluated on a selected digitized dataset compiled from 13 experiments. The relatively larger spread (compared to electrons) is attributed to the larger dataset and data offset of the Ref. [100].

Model performance differences are quantified using the relative residuals for the "All sources" subset. Fig. 7(c) provides insight into the fit quality of both global and selected electric field ranges. Over the full field range (0.009–12 V μm$^{-1}$), all three models exhibit small systematic deviations, with global median signed relative residuals of 2.33% (TK), 2.12% (CT), and 2.17% (PW). The 68% absolute relative-error bounds are very similar, at 8.09% (TK), 8.12% (CT), and 8.17% (PW), while the 95% bounds remain below 18.40%, 18.34%, and 17.91%, respectively.

At low fields ($E < 0.1$ V μm$^{-1}$), the PW model remains nearly unbiased, with median signed errors of −1.79% (TK), −2.94% (CT), and −0.09% (PW). In this regime, the PW model achieves the narrowest error distribution, with 68% of its absolute relative errors below 3.57%, compared to 5.79% and 6.57% for the TK and CT models, respectively. In the mid-field region (0.1–4.0 V μm$^{-1}$), relevant for typical detector operation, the performance of all three models is very similar, with median signed residuals of 2.39% (TK), 2.29% (CT), and 2.45% (PW), and 68% absolute relative-error bounds of 8.12%, 8.13%, and 8.28%, respectively, supporting the use of any of the three parameterizations for hole-drift analysis in this range.

At high fields ($E \geq 4$ V μm$^{-1}$), the TK and CT models show modest positive biases, with median signed errors of 4.65% and 2.57%, respectively. In contrast, the PW model is nearly unbiased, with a median signed error of −1.06%. The 68% absolute relative-error bounds in this regime are 6.87% (TK), 4.44% (CT), and only 1.18% (PW), indicating that the PW parameterization provides the most accurate description of high-field hole drift for the unscaled "All sources" dataset.

These residual-based metrics, together with the global performance indicators in Table 5 (2nd block), show that for the unscaled "All sources" subset the PW model attains the lowest $\chi^2$/ndf (76.33, compared to 78.87 and 78.69 for TK and CT) and substantially lower information criteria (AIC = 38253, BIC = 38274). This provides strong statistical support for the PW model across the full electric-field range of holes when source-dependent normalization offsets are not explicitly corrected.

We also analyzed the "$All_{LA}$" and "$All_{AL}$" for holes. Fig. 7(d) shows the fit quality for the Laser-to-Alpha scaled subset "$All_{LA}$" (Table 5, 2nd block, 4th column). The global median signed residuals are close to 1% for all three models (TK-1.01%, CT-0.86%, and PW-0.92%). The 68% absolute relative-error bounds over full electric field range 0.009–12 V μm$^{-1}$ are reduced to 6.87%, 6.66%, and 6.66% for TK, CT, and PW, respectively. In the mid-field region (0.1–4.0 V μm$^{-1}$), the three parameterizations remain practically indistinguishable, with 68% absolute relative-error bounds around 6.6–6.7%. At high fields ($E \geq 4$ V μm$^{-1}$), the TK model still shows a positive bias (median 5.44%), whereas the CT and PW models are effectively unbiased (median signed errors of 0.25% and 0.02%, with 68% absolute relative errors of 1.80% and 1.52%, respectively).

For the Alpha-to-Laser scaled subset $All_{AL}$ (Table 5, 2nd block, 5th column), the picture is very similar. Global median signed residuals remain below 1% for all models, with 68% absolute relative-error bounds of about 6.7%. Again, CT and PW are nearly indistinguishable in terms of residual distributions, but the CT model achieves slightly lower $\chi^2$/ndf and information criteria in both scaled combinations. In particular, for $All_{LA}$ the CT model attains $\chi^2$/ndf = 50.43 with BIC = 25332, compared to 50.56 and BIC = 25360 for the PW model, and for $All_{AL}$, the corresponding values remain slightly in favor of CT as well.

Source-specific fits further support this picture. For the alpha-only subset, all three models describe the data well, with global median signed residuals at the level of 0.5–1.4% and 68% absolute relative-error bounds around 6.3%, while the CT and PW models yield very similar $\chi^2$/ndf ($\approx$ 48). For the laser-only subset, the residuals are also small, but the statistics are much more limited (74 data points from three experiments with fewer than six samples), compared to the

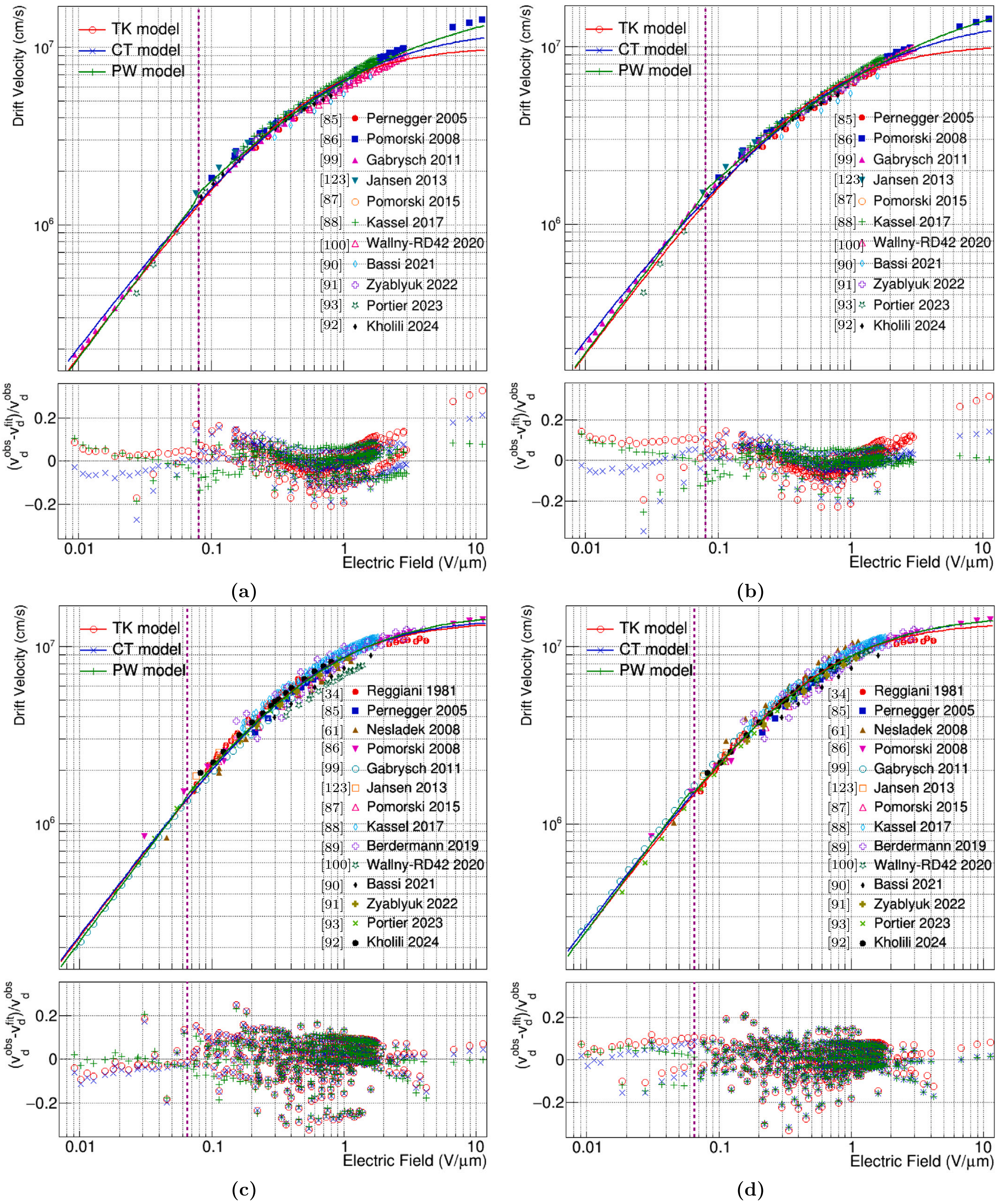

**Fig. 7.** Electron (top) and hole (bottom) $v_d(E)$ data for diamond, compiled from TCT experiments over full electric field range 0.009–12 V µm$^{-1}$ with alpha, laser and electrons as ionizing sources (a, c) "All" category with unscaled data (b, d) "All$_{LA}$" category with laser data scaled to alpha by a factor of 1.091 (electrons) and 1.219. The data are used to fit and evaluate the TK, CT, and PW models (lines), with data shown as markers. Relative residuals $(v_d^{obs} - v_d^{fit})/v_d^{obs}$ versus E are plotted to aid visualization of fit quality.

alpha subset with 369 data points from eight experiments and more than 45 samples investigated. Consequently, the alpha-source data provide the statistically more robust constraint on the hole-transport parameterization.

Taken together, these results indicate that (i) all three models perform comparably well for holes over the full field range once source-dependent normalization is controlled, (ii) in the scaled "All" combinations the CT model consistently attains the lowest AIC/BIC with a simpler functional form. As seen in Table 5, the CT model effectively reduces to the TK model (with $\beta \approx 1$), which explains its early adoption by Reggiani et al. [34] for hole transport in diamond and its subsequent use in later experimental work for both holes and electrons. The nearly similar global metrics of the PW model compared with the TK and CT models suggest that, for holes, the additional scattering mechanism included in the PW parameterization is negligible compared to the repopulation effect discussed for electron drift (see Section 6.1.1).

On this basis, we therefore adopt the CT parameterization as the reference model for hole drift (see Table 11, 2nd row). The extracted parameter values of $v_s$, $E_c$, $\mu_{0,1}$, and $\beta$ for all three models are listed in Table 13 (see Appendix A).

### 5.5.2. Detector operational (DO) field range

The HEP diamond-detector community is primarily interested in model performance within the detector-operation (DO) electric-field range, and many TCT characterizations are therefore performed around this regime. In HEP experiments, detectors typically operate near ≈ 1 V μm$^{-1}$. Although this DO range is contained within the full electric-field interval considered above, dedicated performance metrics in the interval 0.1–4.0 V μm$^{-1}$ are highly relevant for HEP applications. Restricting the analysis to this range suppresses the influence of very low- and high-field data on the global figures of merit and directly tests model behavior where detectors are actually operated. We, therefore, looked at this mid-field in the residual analysis in Section 5.5.1 and re-evaluated all models on the pooled electron and hole datasets restricted to 0.1–4.0 V μm$^{-1}$. The corresponding metrics are summarized in Table 14, (Appendix B).

**Electrons:** Within the DO range, the CT and PW models become practically indistinguishable. For the "All", "Alpha", and "Laser" subsets, both exhibit small residual biases (at the ∼1% level) and similar 68% absolute relative-error bounds (around 5%–6%), clearly improving on the TK model. Thus, in the DO range either CT or PW can be used without loss of accuracy. We retain the PW model as the generalized framework because it also provides an excellent description over the full electric-field range.

**Holes:** For holes, the three models are even more similar in the DO range. Across the "All", "Alpha", and "Laser" subsets, TK, CT, and PW all show small, centrally distributed residuals (biases at the ∼1% level) and comparable 68% absolute relative-error bounds of about 6%–8%. In practice, the models are interchangeable in this field range, consistent with the full–field-range analysis.

## 6. Dependence of mobility on temperature

This section reviews and analyzes experimental work on charge-carrier mobility as a function of temperature. As temperature increases, the phonon population grows, thereby enhancing carrier-phonon scattering (see Section 3). Temperature-dependent characterization of mobilities was first performed using Hall-mobility techniques [63–68,84], and later refined by TCT studies [31–34,61,62,97–99,117,120,123].

Table 6 summarizes electron and hole mobilities versus temperature obtained from Hall-mobility and TCT measurements.

Near room temperature, both electrons and holes follow the trend $\mu \propto T^{-1.5}$. Fig. 8 shows the mobility data from the experiments discussed here. TCT experiments have reported mobility values on the

**Table 6**
Charge-carrier mobilities in diamond as a function of temperature.

| Ref. | T (K) | $\gamma$ (Eq. (6)) | Model[a] |
|---|---|---|---|
| Electrons | | | |
| [63] | 105–300 | $T^{-1.5}$ | (20) |
| [64] | 90–277 | $T^{-1.5}$ | (20) |
| [67] | 200–1000 | $T^{-1.5}$ near 290 K; $T^{-3}$ for $T > 400$ K | (20) |
| [68] | 120–390 | $T^{-1.5}$ | (20) |
| [31] | 85–700 | $T^{-1.5}$ for $T < 400$ K; steeper at higher $T$ | (4) |
| [61] | 320–560 | $T^{-1.4}$ for $300 < T < 340$ K; $T^{-2.5}$ at higher $T$ | (4) |
| [99] | 83–460 | $T^{-1.6\pm0.1}$ for $83 < T < 260$ K; $T^{-2.7\pm0.1}$ at higher $T$ | (4) |
| [62] | 2–295 | – | – |
| Holes | | | |
| [64] | 90–277 | $T^{-1.5}$ | (20) |
| [65] | 100–600 | $T^{-1.5}$ | (20) |
| [66] | 200–800 | $T^{-2.8}$ for $T > 400$ K | (20) |
| [31] | 85–700 | $T^{-1.5}$ for $T < 400$ K; steeper at higher $T$ | (4) |
| [97] | 300–540 | $T^{-1.44}$ for $T < 360$ K; $T^{-3.66}$ at higher $T$ | (28) |
| [98] | 85–470 | $T^{-1.55}$ for $150 < T < 320$ K | (2) |
| [98] | 85–470 | $T^{-3.4}$ for $350 < T < 470$ K | (2) |
| [61] | 273–550 | $T^{-1.5}$ for $300 < T < 340$ K; $T^{-3.2}$ at higher $T$ | (4) |
| [99] | 83–460 | $T^{-1.7\pm0.1}$ | (4) |
| [62] | 2–295 | – | – |

*Note.* Explicit temperature scaling is not provided in Ref. [62], but the data are valuable for reanalysis.

[a] The "Model" column indicates the model equations used to extract mobilities at the given temperatures.

order of $10^6$ cm$^2$/Vs for both electrons [125] and holes [93] at $T \le 10$ K. Electron-mobility data exhibit a broader spread at a given temperature than hole mobility, which can be attributed to the intervalley repopulation effect in the diamond conduction band (see Section 6.1.1).

Time-resolved cyclotron resonance (TRCR) measurements have reported electron mobilities on the order of $10^7$ cm$^2$/Vs. The mobilities extracted by this contactless spectroscopic technique are termed AC mobilities and are given by

$$\mu = \frac{qt_p}{m^*}, \quad \text{where} \quad t_p = \frac{B_0}{\omega\Gamma} \quad \text{and} \quad m^* = \frac{qB_0}{\omega}. \tag{31}$$

Here, $t_p$ is the momentum-relaxation (scattering) time, $B_0$ is the resonance magnetic field, $\omega$ is the microwave angular frequency, and $\Gamma$ is the spectral linewidth [125].

### 6.1. Electrons

TCT measurements across temperatures reveal three distinct regimes of electron transport in diamond (see Fig. 9(a)).

(i) T ≲ 120 K: In this regime, TCT signals show a clear two-step falling edge, indicating two electron populations. Slower carriers with a high longitudinal effective mass ("cool" electrons) and faster carriers with a small transverse effective mass ("hot" electrons) [93] (see Section 6.1.1). At very low electric fields the $v_d$ scales nearly linearly (on a logarithmic scale) up to the field $E_{min}(T)$ (see Section 6.1.2), after which it decreases following a brief saturation as E increases (see Figs. 9(a), 10) i.e., the curve exhibits negative differential mobility $dv/dE < 0$ [127,128].

(ii) T ∼ 120–220 K: The $v_d$–E characteristics display an initial linear trend (at very low electric fields) up to $E_{min}(T)$, where a pseudo-saturation plateau appears that widens and acquires a positive slope with increasing temperature (see Figs. 9(a), 10), followed by a renewed linear region and then a true saturation region.

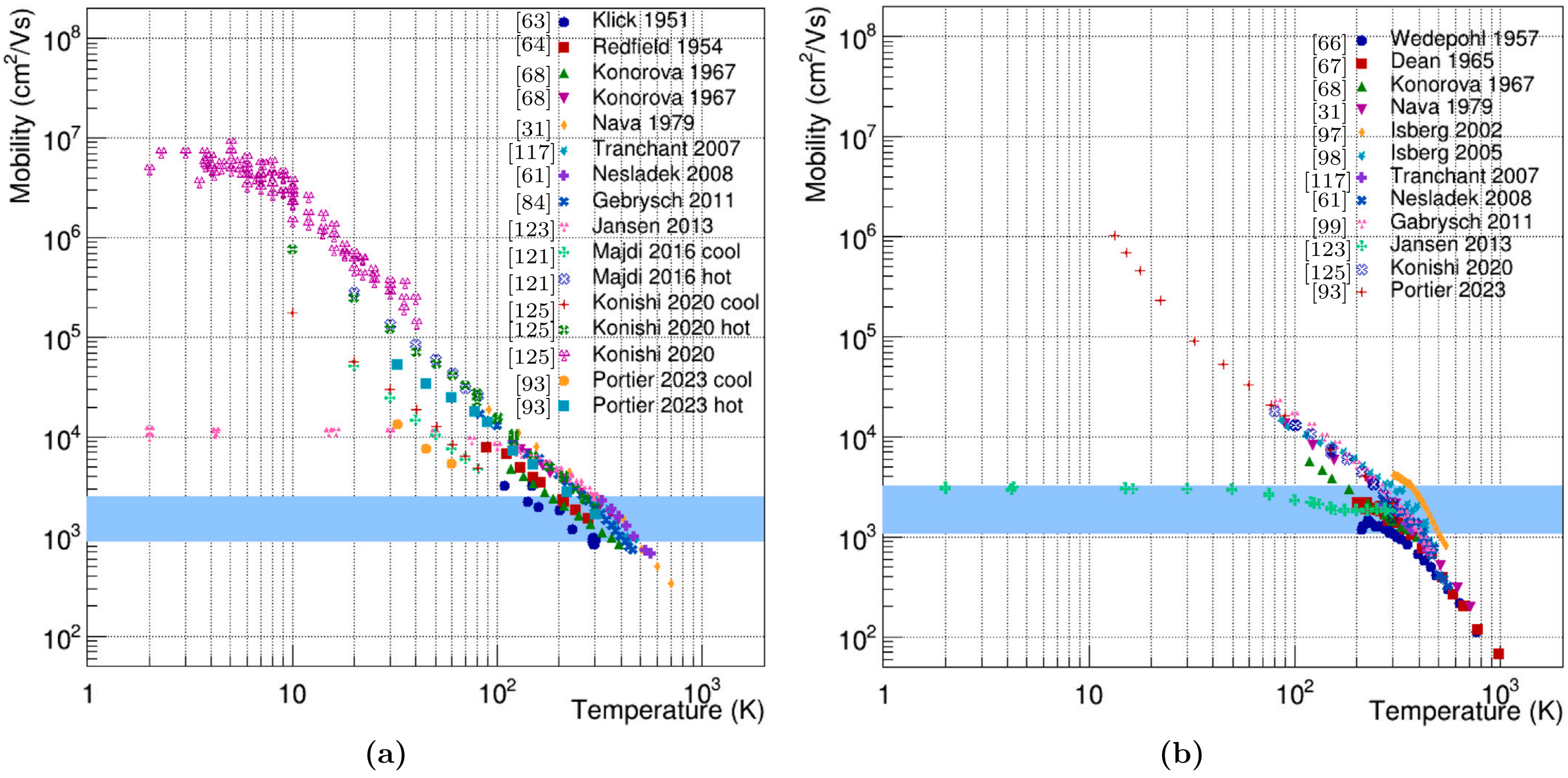

**Fig. 8.** Carrier mobility vs temperature digitized from the literature, (a) electrons (b) holes. Points show individual datasets (see legend). The shaded band denotes the spread of reported 300 K mobilities.

(iii) $T \geq 220$ K: The pseudo-saturation plateau vanishes and the $v_d$-E curve becomes monotonic, with a single (approximately) linear region followed by the onset of saturation at $E_c$. Consequently, the TK, CT, and PW models, each capturing a linear segment followed by non-linear behavior that saturates, are most applicable in this temperature regime. They do not account for negative differential mobility or multi-plateau features evident at lower temperatures.

### *6.1.1. Repopulation effect*

Diamond is a multivalley semiconductor: its conduction band contains six equivalent minima located along the ⟨100⟩ directions (±[100], ±[010], ±[001]). Electron transport involves both intervalley and intravalley scattering among these anisotropic energy valleys. When an electric field is applied, the sixfold degeneracy of the conduction-band minima is lifted, leading to valley polarization. The two valleys aligned with the field, characterized by a larger longitudinal effective mass, are commonly referred to as "cool" valleys, while the four transverse valleys, with lighter transverse effective mass, are termed "hot" valleys owing to their enhanced energy gain under acceleration.

Intervalley-scattering probability from "cool" to "hot" valleys is high at low electric fields, which results in a larger population of electrons in the hot valleys. However, as the field strength increases, this trend reverses and most carriers repopulate the cool valleys. This redistribution of carriers, known as the Repopulation Effect, was first observed through Hall-mobility measurements in diamond by Lenz in 1927 [63,129]. Because repopulation significantly affects electron transport in diamond, quantitative models must account for the underlying valley-polarized transport. In this section, we adopt a superposition framework wherein the total $v_d$ is modeled as the population-weighted sum of the hot- and cool-valley contributions, with population fractions determined by field- and temperature-dependent intervalley-scattering rates.

The application of a magnetic field further lifts the residual degeneracy of the transverse (hot) valleys, inducing a secondary redistribution of carriers and a further reduction in drift velocity [130]. This **secondary repopulation effect** is particularly relevant for diamond-based HEP detectors operating in strong magnetic fields and merits deeper investigation.

### *6.1.2. Superposition model*

Electron transport in diamond can be more accurately modeled by assuming a superposition of two distinct drift-velocity states, one associated with carriers in the hot valleys and the other with carriers in the cool valleys. The total drift velocity, $v_d(E,T)$, is then expressed as,

$$v_d(E,T) = P_H(E,T) \cdot v_{d,H}(E,T) + P_C(E,T) \cdot v_{d,C}(E,T). \tag{32}$$

where $v_{d,H}$ and $v_{d,C}$ denote the drift velocities in the hot and cool valleys, respectively, and $P_H(E,T)$, $P_C(E,T)$ are their corresponding occupation probabilities. The probabilities satisfy the normalization condition: $P_H + P_C = 1$.

As shown in Fig. 9(a), the $v_d - E$ characteristics between 100–240 K exhibit a pseudo-saturation plateau. The onset field, slope, and width of this plateau are strongly temperature-dependent (see Fig. 10). The initial and final points of the plateaus are listed in Table 7, each column is fit linearly to obtain the empirical relationships given in Eqs. (35) and (36).

The difference $\Delta E$, defined in Eq. (34), characterizes the plateau width and its evolution with temperature. Although at higher temperatures the plateau becomes increasingly obscured in $v_d$ curves due to enhanced phonon scattering, extrapolation based on $\Delta E(T)$ allows estimation of the effective repopulation region.

This extrapolated plateau region is particularly significant within the superposition framework, as it marks the field interval over which the valley-occupation probabilities $P_H(E,T)$ and $P_C(E,T)$ undergo their most rapid transitions. This transition governs the repopulation effect and strongly influences the overall drift velocity. For modeling purposes, $P_C(E,T)$ is approximated as a linearly varying "cross-fade" function across this interval.

$$P_C(E,T) = \begin{cases} 0, & E \leq E_{\min}(T), \\ \dfrac{E - E_{\min}(T)}{\Delta E(T)}, & E_{\min}(T) < E < E_{\max}(T), \\ 1, & E \geq E_{\max}(T), \end{cases} \tag{33}$$

such that $P_H = 1 - P_C$. The width of the plateau $\Delta E(T)$ is expressed as

$$\Delta E(T) = E_{\max}(T) - E_{\min}(T). \tag{34}$$

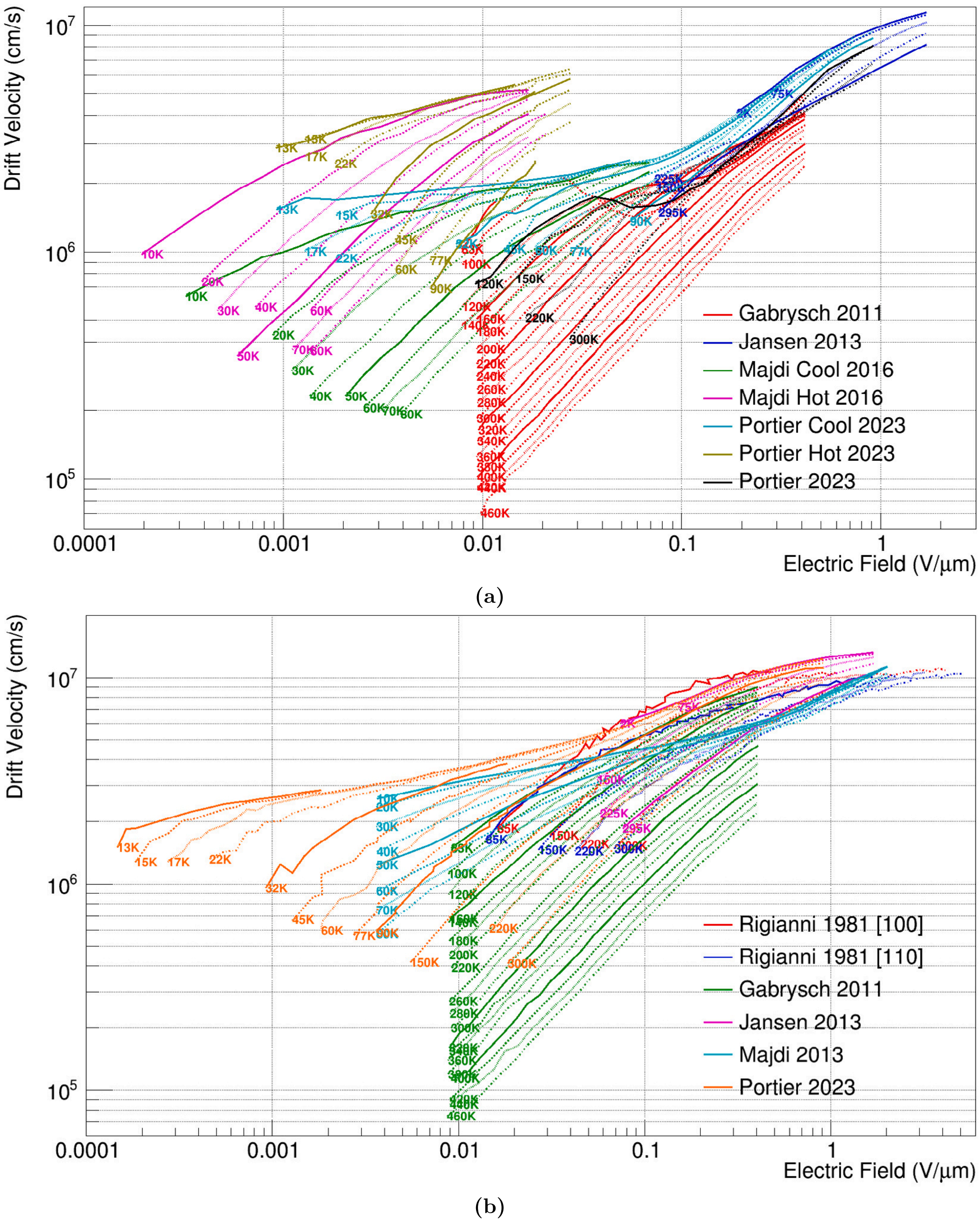

**Fig. 9.** Drift velocity $v_d(E,T)$ versus electric field and temperature, obtained by the TCT technique using laser excitation [99], $^{241}$Am [62], and electron-beam sources [34,93]. (a) *Electrons:* Portier (2023) hot/cool (olive)/(cyan), $T = \{13, 15, 17, 22, 32, 45, 60, 77, 90\}$ K [93], Majdi (2016) hot/cool (magenta)/(green), $T = 1080$ K (10 K steps) [121], Portier (2023) (black), $T = \{120, 150, 220, 300\}$ K [93], Jansen (2013) (blue), $T = \{2, 75, 150, 225, 295\}$ K [62], Gabrysch (2011) (red), $T = 83$ K and 100460 K (20 K steps) [99]. (b) *Holes:* Portier (2023) (orange), $T = \{13, 15, 17, 22, 32, 45, 60, 77, 90, 150, 220, 300\}$ K [93], Majdi (2013) (cyan), $T = 1080$ K (10 K steps) [126], Jansen (2013) (magenta), $T = \{2, 75, 150, 225, 295\}$ K [62], Gabrysch (2011) (green), $T = 83$ K and 100460 K (20 K steps) [99], Reggiani (1981) (red)/(blue) for [100]/[110], $T = \{85, 150, 220, 300\}$ K [34]. Traces in each panel are ordered left-to-right by increasing $T$. Colors correspond to the legend in the main text.

The onset and termination fields of the $v_d$ plateau, $E_{min/max}$, exhibit linear temperature dependence and are fitted using data from [62,93,99], shown in Table 7.

$$E_{min}(T) \approx -5.57816 \times 10^{-2} + 7.00760 \times 10^{-4}\, T, \tag{35}$$

$$E_{max}(T) \approx -1.5732869 \times 10^{-1} + 2.04800 \times 10^{-3}\, T. \tag{36}$$

Eqs. (35) and (36) allow estimation of the plateau width as a function of temperature. These relationships are used to extrapolate the plateau region at 300 K, where the effect is no longer directly observable due to phonon-scattering suppression. The extrapolated field range is then applied to TCT data for electron drift at 300 K, as shown in Fig. 11(b).

At 300 K, the superposition model (Eq. (32)) can be approximated by the PW model under the assumption that $v_d$ saturation does not occur in the hot valleys. Consequently, $P_H$ goes to zero before the onset of hot-valley saturation, resulting in a computationally efficient formulation that preserves the essential underlying charge-carrier transport physics. In this sense, the PW model can be expressed as a combination of the Drude model (Eq. (2)) and the CT model (Eq. (5)), i.e., H(DM)+C(CT), where DM denotes the Drude model.

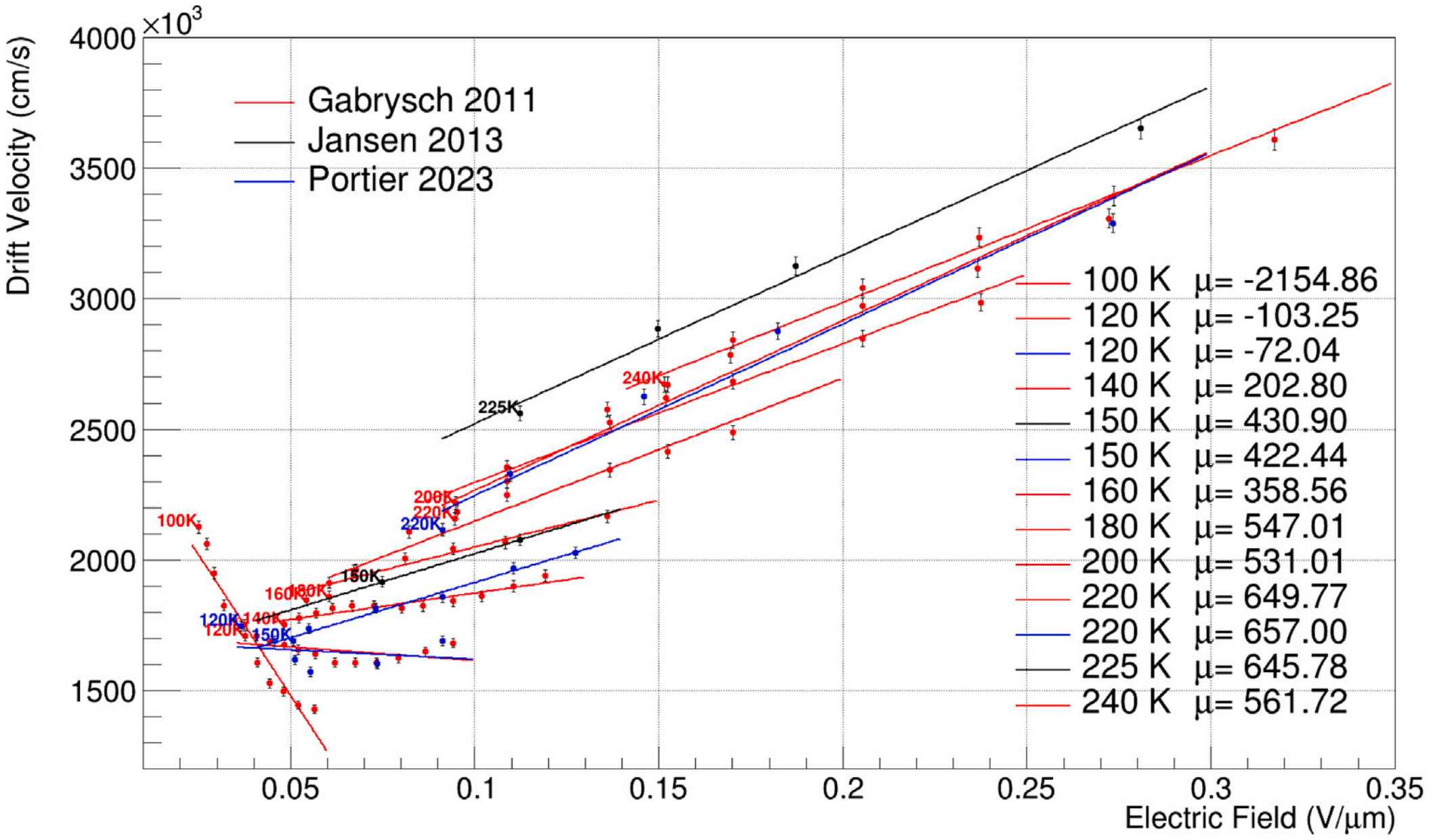

**Fig. 10.** Drift velocity $v_d$ versus electric field E plateau region. Data are taken from three TCT experiments using a laser [99], Am-241 [123], and electrons [93] as the ionizing sources. Here $\mu = v_d/E$, showing the negative mobility that gradually recovers with increasing temperature.

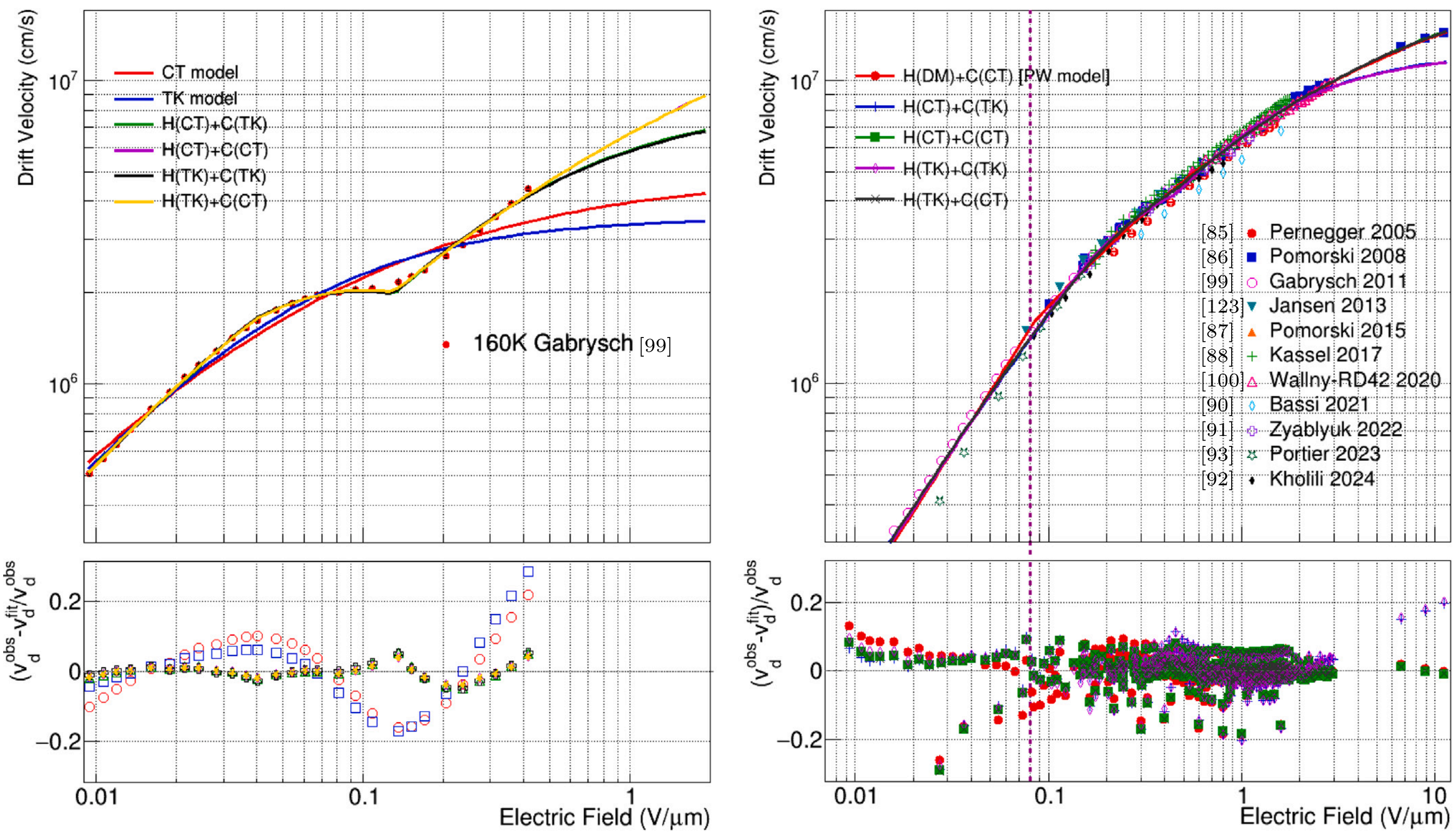

(a) (b)

**Fig. 11.** Evaluation of TK, CT, PW, and superposition (Eq. (32)) models using $v_d$–$E$ data from (a) Ref. [99] at 160 K showing best performance of superposition framework and (b) 11 TCT experiments at room temperature (laser data scaled to alpha), showing the validation of approximation of superposition model to PW model at room temperature.

**Table 7**
Electric-field plateau (dominant intervalley scattering: hot → cool valleys) versus temperature.

| $T$ (K) | $E_{min}$ (V/μm) | $E_{max}$ (V/μm) |
|---|---|---|
| 100 | 0.023 | 0.060 |
| 120 | 0.035 | 0.100 |
| 140 | 0.045 | 0.130 |
| 150 | 0.040 | 0.140 |
| 160 | 0.050 | 0.150 |
| 180 | 0.060 | 0.200 |
| 200 | 0.085 | 0.250 |
| 220 | 0.090 | 0.300 |
| 225 | 0.090 | 0.300 |
| 240 | 0.140 | 0.350 |

**Table 8**
Performance metrics for TK, CT, and superposition combinations obtained by fitting Gabrysch et al. [99] $v_{d(E)}$ data at 160 K.

| Model | $\chi^2$/ndf | $R^2$ | $\sigma_z$ | AIC | BIC |
|---|---|---|---|---|---|
| CT | 104.99 | 0.924 | 9.81 | 2631 | 2635 |
| TK | 112.97 | 0.883 | 10.38 | 2941 | 2944 |
| H(CT)+C(TK) | 5.78 | 0.996 | 2.22 | 143 | 150 |
| H(CT)+C(CT) | 3.94 | 0.997 | 1.79 | 99 | 107 |
| H(TK)+C(TK) | 5.73 | 0.996 | 2.26 | 145 | 151 |
| H(TK)+C(CT) | 3.77 | 0.997 | 1.79 | 97 | 103 |

**Table 9**
Performance metrics for PW model, expressed here as H(DM)+C(CT), and superposition combinations obtained by fitting data (11 experiments) at room temperature.

| Model | $\chi^2$/ndf | $R^2$ | $\sigma_z$ | AIC | BIC |
|---|---|---|---|---|---|
| H(DM)+C(CT) | 21.04 | 0.992 | 4.56 | 8153 | 8172 |
| H(CT)+C(TK) | 23.04 | 0.983 | 4.77 | 8903 | 8923 |
| H(CT)+C(CT) | 18.65 | 0.992 | 4.29 | 7191 | 7215 |
| H(TK)+C(TK) | 23.45 | 0.983 | 4.82 | 9084 | 9100 |
| H(TK)+C(CT) | 18.61 | 0.992 | 4.29 | 7193 | 7213 |

Fig. 11(a) compares the TK, CT, and superposition variants at 160 K and shows that the two-component superposition framework, by explicitly capturing hot–cool valley repopulation, reproduces the pseudo-saturation plateau and achieves order-of-magnitude improvements in fit quality (e.g., $\chi^2$/ndf $\approx$ 3–6 vs. $\gtrsim$ 100 for TK/CT) with substantially lower AIC/BIC (Table 8). Thus, it provides the most faithful descriptive and predictive model of low-temperature electron transport in diamond.

Fig. 11(b) evaluates the PW model (Eq. (30))—expressed as H(DM) +C(CT)—against superposition variants (Eq. (32)) using data from 11 room-temperature TCT experiments. In terms of goodness of fit, the PW model performs comparably to the best two-component superposition models (e.g., $\chi^2/\text{ndf} = 29.38$ vs. 27.18 and $R^2 = 0.981$ vs. 0.982), while using fewer parameters (Table 9). As expected, information criteria (AIC/BIC) tend to favor the more flexible superposition formulations. Nevertheless, the marginal improvement in $\chi^2$/ndf is small over the room-temperature range. Therefore, for practical room-temperature applications, the PW model offers a parsimonious, simulation-friendly approximation to the superposition framework.

### 6.2. Holes

Hole transport as a function of $E$ and $T$ reveals two regimes (see Fig. 9(b)).

(i) $T \lesssim 80$ K: Within the low-field range, $v_d$ shows non-ohmic behavior. The $v_d$–$E$ characteristics in this regime resemble the cool-valley transport of electrons, suggesting the involvement of heavy holes. Majdi et al. [126] propose a heavy-hole–only transport model exhibiting API and nonpolar OPI that describes the experimental data well.

**Table 10**
Temperature scaling of the parameters for TK, CT, and PW models using data from Gabrysch (2011) [99] in the electric-field range 0.009–0.41 V/μm and temperature ranges 220–320 K (electrons) and 160–320 K (holes).

| | | $v_s$ A(cm/s)·$T^\gamma$ | $E_c$ A(V/μm)·$T^\gamma$ | $\beta$ A·$T^\gamma$ |
|---|---|---|---|---|
| (4) | e | $5.74\times10^3\cdot T^{1.26}$ | $1.77\times10^{-9}\cdot T^{3.37}$ | – |
| | h | $2.39\times10^6\cdot T^{0.27}$ | $2.13\times10^{-6}\cdot T^{2.17}$ | – |
| (5) | e | $1.91\times10^3\cdot T^{1.44}$ | $1.16\times10^{-9}\cdot T^{3.43}$ | $4.18\cdot T^{-0.24}$ |
| | h | $4.04\times10^{10}\cdot T^{-1.48}$ | $2\times10^{-3}\cdot T^{0.92}$ | $8.1\times10^{-4}\cdot T^{1.28}$ |

ii) $T \gtrsim 80$ K: The $v_d$–$E$ curves become smooth, with a single linear region followed by saturation, rendering the TK and CT models applicable to this regime for the same reasons discussed in Section 6.1.

### 6.3. Parameter scaling with temperature

In many simulation frameworks (e.g., Allpix-Squared), mobility models describe $v_d(E,T)$ rather than only $v_d(E)$. For example, the Jacoboni–Canali [60] model is widely used for silicon charge-carrier drift across low to high electric fields and temperatures (see Section 3.3.2). A similar approach is needed for diamond.

For electron drift, a practical starting point is regime iii (Section 6.1), where the TK and CT models apply and their parameters ($v_s, E_c, \beta$) can be scaled with temperature using Eq. (6). However, unlike silicon, diamond exhibits multi-step scaling (see Table 6). Owing to limited literature coverage in both electric-field and temperature, we performed temperature scaling over 0.009–0.4 V μm$^{-1}$ and 220–460 K using data from [99] and report a summary in Table 10.

For holes, scaling is performed over 160–320 K and 0.009–0.41 V μm$^{-1}$. Results are given in Table 10. This scaling has limitations. Only one TCT experiment [99] provides measurements between 80–460 K with a temperature step $\Delta T = 20$ K suitable for scaling. Moreover, those data span very-low to moderate-low electric fields (0.009–0.4 V μm$^{-1}$) with no coverage at higher fields. As a result, the extracted saturation velocities $v_s$ for both the TK and CT models are likely underestimated. More accurate parameterizations of A and $\gamma$ will require TCT measurements extending to higher fields on the order of 10 V μm$^{-1}$.

## 7. Conclusion

The large dispersion in reported electron and hole mobilities and saturation velocities in diamond arises primarily from differences in experimental conditions—most notably the electric-field window probed in TCT measurements—and from the analysis model employed. The excitation source (e.g., $\alpha$, laser, electron) further modulates the inferred mobilities.

Historically, the TK model has been used in most studies, owing to the short field windows typical of TCT data and the precedent set by Reggiani et al. [34]. The CT model appears in only a few works, however, our results show that the CT model outperforms the TK model when applied experiment-wise (Table 4). In the Electron Group pooled dataset with a broad field range (0.009–12 V μm$^{-1}$), our proposed PW model (Eq. (30)) provides the best description and, within the DO range (0.1–4.0 V μm$^{-1}$), effectively collapses to CT across all three subsets (see Table 14), in the full-range analysis it also reduces to CT for the “Alpha source” subset (see Table 5). For the Hole Group pooled data, both CT and PW reduce to TK (Tables 5, 14). Nevertheless, the PW (for electrons) and CT (for holes) models remain convenient frameworks for implementation in simulation tools.

For practical comparison across excitation sources and crystal qualities, we normalize all drift-velocity data to a common “$\alpha$-scale” defined by the combined set of $\alpha$-TCT measurements, rather than by a single ultra-pure reference sample. For each carrier type, in the overlapping electric-field range between the $\alpha$ and laser–TCT datasets we determine a global multiplicative factor $x$ such that

$$v_{d,\alpha}(E) \simeq x\, v_{d,\text{laser}}(E),$$

with $x$ obtained from an unweighted least-squares regression constrained through the origin and cross-checked by the median of the pointwise ratios $v_{d,\alpha}/v_{d,\text{laser}}$. This procedure brings all curves onto the same $\alpha$-based scale. For electrons, a single factor describes the offset at the few-percent level, whereas for holes it provides only an approximate normalization with residual variations at the $\gtrsim$10% level.

**Table 11**
Parameters of the best performing models (Bold in Table 5) in the full electric field range (0.009–12 V/μm), that we recommend for use in simulation frameworks.

| Carrier | Model | $v_s$ ($10^6$ cm/s) | $E_c$ (V/μm) | $\mu_{0,1}$ ($cm^2$/Vs) | $\mu_{0,2} = v_s/E_c$ ($cm^2$/Vs) | $\beta$ (–) |
|---|---|---|---|---|---|---|
| Electrons | PW | 26.59 ± 0.51 | 0.56 ± 0.00 | 1880 ± 4 | 4748 ± 92 | 0.41 ± 0.00 |
| Holes | CT | 15.09 ± 0.05 | 0.55 ± 0.00 | – | 2744 ± 9.4 | 0.88 ± 0.00 |

The upper envelope of the $\alpha$ measurements, dominated by high-quality scCVD crystals such as Pomorski (2008), that provides an approximate reference as a near-ideal material. $\alpha$ data taken on lower-quality crystals (e.g. with shorter carrier lifetimes, as in Pernegger or Bassi samples) tend to lie on or even below the laser-based curves, and their offset from the upper envelope can thus be used as a practical indicator of reduced material quality. The resulting scale factors should therefore be interpreted as effective corrections that absorb both source-dependent systematics and sample-to-sample variations, rather than as a strict calibration to a unique “absolute” ultra-pure reference. Additional high-quality drift-velocity measurements, together with impurity and defect characterization, will be required to quantitatively define this scale and to disentangle remaining source and material effects.

Electron drift is additionally influenced by a repopulation effect that is most apparent at low temperature in high-quality material, where phonon scattering is suppressed. We observe an electric-field plateau in $v_d(E)$ whose slope and extent are strongly temperature dependent. At very low temperatures ($T \lesssim 120$ K) the slope becomes negative—consistent with negative differential mobility (Fig. 10)—whereas with increasing temperature the slope turns positive and the plateau length grows, reflecting competition between intervalley scattering and phonon scattering.

To capture this behavior across wide electric-field and temperature ranges, we introduced a superposition model that explicitly incorporates repopulation. It reproduces electron drift from cryogenic to room and elevated temperatures, whereas TK, CT, and PW reliably describe the data only above about 220 K. In the limit where electrons preferentially scatter into cool valleys before hot-valley saturation, the superposition model reduces to the PW model, depending on the available field window, further simplifications recover CT or TK. A comparative summary for electrons and holes, including source dependence, is given in Tables 5, 14 and 9.

Finally, we provide temperature scalings for the TK and CT models (Table 10) over a narrow interval (electrons: 220–320 K, holes: 160–320 K) around room temperature. Because the models’ parameters exhibit stepwise temperature dependence and existing measurements are insufficient to map those steps comprehensively, we adopt a single scaling interval that includes room temperature.

**Implications for simulation.** For practical device and detector simulations, our statistical comparison indicates that the piecewise (PW) mobility model provides the most consistent description for electrons and the Caughy Thomas (CT) mobility model for holes over broad electric-field ranges at (and near) room temperature. We therefore recommend using the parameter set in the Table 11 as generalized values for implementation in simulation frameworks. These values can be scaled down by a simple multiplicative factor to adjust with source types, flux rates and impurity/defects concentrations.

**Outlook.** Tighter parameter estimates and robust temperature scalings require TCT measurements on the same detector samples with (i) electric-field coverage from below $1 \times 10^{-4}\,\text{V}\,\mu\text{m}^{-1}$ to above $10\,\text{V}\,\mu\text{m}^{-1}$, and (ii) temperature coverage from tens of K up to ~1000 K with steps of order 20 K. At present, [99] provides a valuable but low-field dataset over 83–460 K (~20 K steps), which leaves $v_{sat}$ weakly constrained and obscures the temperatures at which parameter scalings change. A single, consistent TCT campaign over wide (E, T) would reduce these uncertainties and enable global fits for both electrons and holes.

## CRediT authorship contribution statement

**Faiz Rahman Is-haqzai:** Writing – review & editing, Writing – original draft, Visualization, Validation, Software, Methodology, Investigation, Formal analysis, Data curation, Conceptualization. **Muhammed Deniz:** Writing – review & editing, Visualization, Supervision, Resources, Project administration. **Kevin Kröninger:** Writing – review & editing, Validation, Supervision, Resources, Project administration, Funding acquisition. **Jens Weingarten:** Writing – review & editing, Visualization, Supervision, Project administration, Conceptualization.

## Declaration of Generative AI and AI-assisted technologies in the writing process

During the preparation of this work the authors used ChatGPT and Grammarly in order to improve wording and sentence structure. After using these tools, the authors reviewed and edited the content as needed and take full responsibility for the content of the publication.

## Declaration of competing interest

The authors declare the following financial interests/personal relationships which may be considered as potential competing interests: Faiz ur rahman ishaqzai reports financial support was provided by Alexander von Humboldt Foundation. If there are other authors, they declare that they have no known competing financial interests or personal relationships that could have appeared to influence the work reported in this paper.

## Acknowledgments/funding

This work was supported by funds from the Alexander von Humboldt Foundation.

## Appendix A. Parameters of TK, CT, and PW models extracted by the analysis of Electrons and Holes pooled data

See Tables 12 and 13.

## Appendix B. Performance metrics of TK, CT, and PW models at DO range

See Table 14.

**Table 12**
Parameters in the full electric field range (0.009–12 V/μm). (Electron Group).

| Model | Dataset | $v_s$ ($10^6$ cm/s) | $E_c$ (V/μm) | $\mu_{0,1}$ ($cm^2$/Vs) | $\mu_{0,2} = v_s/E_c$ ($cm^2$/Vs) | $\beta$ (–) |
|---|---|---|---|---|---|---|
| TK | All | 10.11 ± 0.01 | 0.55 ± 0.00 | – | 1838 ± 5 | – |
| | $All_{LA}$ | 10.25 ± 0.01 | 0.55 ± 0.00 | – | 1865 ± 2 | – |
| | $All_{AL}$ | 9.39 ± 0.01 | 0.55 ± 0.00 | – | 1706 ± 2 | – |
| | Alpha | 10.49 ± 0.02 | 0.58 ± 0.00 | – | 1797 ± 7 | – |
| | Laser | 9.33 ± 0.02 | 0.50 ± 0.00 | – | 1870 ± 9 | – |
| CT | All | 13.01 ± 0.07 | 0.60 ± 0.00 | – | 2157 ± 14 | 0.75 ± 0.00 |
| | $All_{LA}$ | 15.21 ± 0.09 | 0.63 ± 0.00 | – | 2414 ± 15 | 0.66 ± 0.00 |
| | $All_{AL}$ | 13.96 ± 0.09 | 0.63 ± 0.00 | – | 2216 ± 14 | 0.65 ± 0.00 |
| | Alpha | 30.71 ± 0.81 | 0.55 ± 0.00 | – | 5571 ± 155 | 0.38 ± 0.01 |
| | Laser | 12.60 ± 0.13 | 0.59 ± 0.00 | – | 2132 ± 27 | 0.72 ± 0.01 |
| PW | All | 20.64 ± 0.31 | 0.54 ± 0.00 | 1771 ± 4 | 3835 ± 62 | 0.47 ± 0.01 |
| | $All_{LA}$ | 26.59 ± 0.51 | 0.56 ± 0.00 | 1880 ± 4 | 4748 ± 92 | 0.41 ± 0.00 |
| | $All_{AL}$ | 24.45 ± 0.47 | 0.56 ± 0.00 | 1722 ± 4 | 4366 ± 85 | 0.41 ± 0.00 |
| | Alpha | 31.11 ± 0.82 | 0.55 ± 0.00 | 1946 ± 19 | 5667 ± 157 | 0.37 ± 0.01 |
| | Laser | 18.98 ± 0.63 | 0.55 ± 0.01 | 1815 ± 5 | 3478 ± 123 | 0.48 ± 0.01 |

**Table 13**
Parameters in the full electric field range (0.009–12/μm). (Holes Group).

| Model | Dataset | $v_s$ ($10^6$ cm/s) | $E_c$ (V/μm) | $\mu_{0,1}$ ($cm^2$/Vs) | $\mu_{0,2} = v_s/E_c$ ($cm^2$/Vs) | $\beta$ (–) |
|---|---|---|---|---|---|---|
| TK | All | 13.92 ± 0.02 | 0.60 ± 0.00 | – | 2320 ± 3 | – |
| | $All_{LA}$ | 13.70 ± 0.02 | 0.54 ± 0.00 | – | 2536 ± 3 | – |
| | $All_{AL}$ | 11.13 ± 0.01 | 0.54 ± 0.00 | – | 2062 ± 2 | – |
| | Alpha | 14.25 ± 0.02 | 0.60 ± 0.00 | – | 2390 ± 8 | – |
| | Laser | 10.97 ± 0.02 | 0.48 ± 0.00 | – | 2306 ± 10 | – |
| CT | All | 14.46 ± 0.05 | 0.61 ± 0.00 | – | 2370 ± 8 | 0.95 ± 0.00 |
| | $All_{LA}$ | 15.09 ± 0.05 | 0.55 ± 0.00 | – | 2744 ± 9 | 0.88 ± 0.00 |
| | $All_{AL}$ | 12.30 ± 0.04 | 0.55 ± 0.00 | – | 2237 ± 8 | 0.87 ± 0.00 |
| | Alpha | 16.46 ± 0.10 | 0.59 ± 0.00 | – | 2786 ± 19 | 0.81 ± 0.01 |
| | Laser | 11.11 ± 0.06 | 0.48 ± 0.00 | – | 2322 ± 15 | 0.98 ± 0.01 |
| PW | All | 15.77 ± 0.08 | 0.59 ± 0.00 | 2155 ± 4 | 2674 ± 13 | 0.83 ± 0.00 |
| | $All_{LA}$ | 15.16 ± 0.07 | 0.56 ± 0.00 | 2492 ± 5 | 2708 ± 12 | 0.87 ± 0.00 |
| | $All_{AL}$ | 12.32 ± 0.05 | 0.55 ± 0.00 | 2038 ± 4 | 2240 ± 10 | 0.87 ± 0.00 |
| | Alpha | 15.83 ± 0.09 | 0.60 ± 0.00 | 2419 ± 11 | 2623 ± 18 | 0.86 ± 0.01 |
| | Laser | 12.81 ± 0.12 | 0.43 ± 0.00 | 2098 ± 4 | 2946 ± 34 | 0.76 ± 0.01 |

**Table 14**
Model performance for drift-velocity fits to TCT data, comparing the TK, CT, and PW models at DO range.

| Carrier E-Range | Model | All | | $All_{LA}$ | | $All_{AL}$ | | Alpha | | Laser | |
|---|---|---|---|---|---|---|---|---|---|---|---|
| | | $\chi^2$/ndf | BIC | $\chi^2$/ndf | BIC | $\chi^2$/ndf | BIC | $\chi^2$/ndf | BIC | $\chi^2$/ndf | BIC |
| Electrons | TK | 41.13 | 14 860 | 36.29 | 13 111 | 36.40 | 13 151 | 29.63 | 8724 | 27.70 | 1614 |
| (0.1–4.0) | CT | 27.08 | 9765 | 16.05 | 5794 | 16.03 | 5789 | 13.10 | 4732 | 2.91 | 178 |
| (V/μm) | PW | 27.15 | 9777 | 16.09 | 5806 | 16.08 | 5801 | 13.13 | 4744 | 2.96 | 186 |
| Holes | TK | 80.68 | 37 448 | 51.62 | 23 963 | 52.27 | 24 266 | 49.12 | 17 645 | 59.30 | 3151 |
| (0.1–4.0) | CT | 78.89 | 36 543 | 50.61 | 23 450 | 51.29 | 23 764 | 48.26 | 17 293 | 59.61 | 3112 |
| (V/μm) | PW | 79.06 | 36 556 | 50.72 | 23 462 | 51.40 | 23 776 | 48.39 | 17 305 | 60.78 | 3120 |

## Data availability

Data used in this study are available in the TUDOdata research data repository of TU Dortmund University at https://doi.org/10.17877/TUDODATA-2025-MIESBVSN.

## Further reading